\documentclass[reprint,
 amsmath,amssymb,
 aps,
 prl,
]{revtex4-2}

\usepackage{verbatim} 
\usepackage{graphicx}
\usepackage{dcolumn}
\usepackage{booktabs}
\usepackage{xr}
\usepackage{hyperref}
\usepackage{bm}
\usepackage{siunitx} 
\usepackage[utf8]{inputenc}
\usepackage[T1]{fontenc} 
\newcommand{\ro}{r_{0}}

\makeatletter
\renewcommand\frontmatter@abstractwidth{\dimexpr\textwidth\relax}
\makeatother

\begin{document}

\preprint{APS}

\title{Tracking the nonlinear formation of an interfacial wave spectral cascade \\ from one to few to many}

\author{Sean M. D. Gregory$^{1}$}%
\email{sean.gregory@nottingham.ac.uk}
\author{Silvia Schiattarella$^{2}$}
\email{silvia.schiattarella@nottingham.ac.uk}
\author{Vitor S. Barroso$^{1}$}%
\author{David I. Kaiser$^{3}$}
\author{Anastasios Avgoustidis$^{2}$}
\author{Silke Weinfurtner$^{1,2}$}
\email{silke.weinfurtner@nottingham.ac.uk}
\affiliation{%
 $^{1}$School of Mathematical Sciences, University of Nottingham, University Park, Nottingham, NG7 2RD, UK
}%
\affiliation{%
 $^{2}$School of Physics and Astronomy, University of Nottingham, University Park, Nottingham, NG7 2RD, UK
}%
\affiliation{%
 $^{3}$Department of Physics, Massachusetts Institute of Technology, Cambridge, MA 02139 USA
}%
\date{\today}

\begin{abstract}

A hallmark of far-from-equilibrium systems is the emergence of a spectral cascade, where energy is transferred across length-scales following a simple power law.
The universal nature of this phenomenon has led to advances in a range of disciplines, including climate forecasting~\cite{lovejoy2018spectra}, foreign exchange trading~\cite{ghashghaie1996turbulent}, and the modelling of neurological activity~\cite{sheremet2019wave}.
For many diverse far-from-equilibrium scenarios, the scaling laws of steady states have been successfully predicted by the statistical theory of weak wave turbulence, originally developed by considering the leading order interactions between waves on a fluid surface~\cite{zakharov2012kolmogorov}.
However, the predictive power of weak wave turbulence breaks down in the presence of large amplitudes, high dissipation, and finite-size effects~\cite{FalconReview22}. 
We offer new insight into these regimes by experimentally tracking the formation of a spectral cascade under these conditions in an externally driven fluid-fluid interface. We resolve individual wave modes and observe their time evolution from one to few to many, a process culminating in a steady state with a spectral density characterised by a power-law scaling.
Our findings confirm that interfacial dynamics can be effectively modelled by a weakly nonlinear Lagrangian theory~\cite{miles1986NonlinearStratified,Barroso_2023}, a predictive framework encompassing both underlying wave interaction and emergent behaviours of the system.
Such nonlinear interactions are experimentally quantified through statistical correlations, revealing a hierarchy in wave-mixing order that confirms a key assumption of weak wave turbulence~\cite{zakharov2012kolmogorov,nazarenko2011wave}.
The Lagrangian formulation further aids our time-evolution analysis; specific interactions are tracked through time, and we predict the timescale until a cascade emerges.  
Our findings are transferable to other far-from-equilibrium systems, which we demonstrate by providing a mapping to reheating scenarios following cosmic inflation in the early Universe~\cite{kofman1997towards,amin2015nonperturbative}.
\end{abstract}

\maketitle

\renewcommand{\figurename}{\textbf{Fig.}}
\renewcommand{\thefigure}{\textbf{\arabic{figure}}}
Spectral cascades have been observed in ocean waves~\cite{nazarenko2016review,FalconReview22}, plasmas~\cite{yamada2008anatomy}, atmospheric flows~\cite{wyngaard1992atmospheric}, superfluid helium~\cite{abdurakhimov2010observation,abdurakhimov2011classical,abdurakhimov2012turbulent}, ultra-cold atoms~\cite{navon2016emergence,navon2019synthetic,zhang2021many,glidden2021bidirectional,Ga_ka_2022,dogra2023universal}, and polaritons~\cite{KONIAKHIN2020109574,baker2023turbulent}.
A notable example is the celebrated Kolmogorov–Zakharov spectrum~\cite{zakharov2012kolmogorov,nazarenko2011wave}, where energy is distributed from an injection range (at large length scales) to a viscous dissipation range (at small scales). While steady-state limits of ``One to Many'' transitions are well understood through weak wave turbulence, the intermediate states remain largely unexplored. 
Understanding this ``few-mode'' regime, characterised by a finite number of interacting modes, is crucial for a complete, deterministic picture of cascade formation; bridging the gap between single-mode coherent dynamics and fully developed out-of-equilibrium states.

Here, we observe the formation of a spectral cascade in a strongly forced, dissipative, and highly discretised regime beyond the applicability of weak wave turbulence theory~\cite{nazarenko2011wave}.
We generate \emph{one} exponentially growing long-wavelength interfacial wave and observe its subsequent decay into a \emph{few} other modes. With time and space-resolved measurements of the interface, we systematically track the individual evolution of 
excited  
waves. From the \emph{few} interactions at early times, we observe the transition of the driven interface into a strongly nonlinear steady-state scaling of \emph{many} interacting waves, reminiscent of wave turbulence~\cite{FalconReview22}. 
We show that the observed far-from-equilibrium dynamics is captured by a finite number of interaction terms in an effective Lagrangian.

\begin{figure*}[ht!]
\centering
\includegraphics[width=\linewidth]{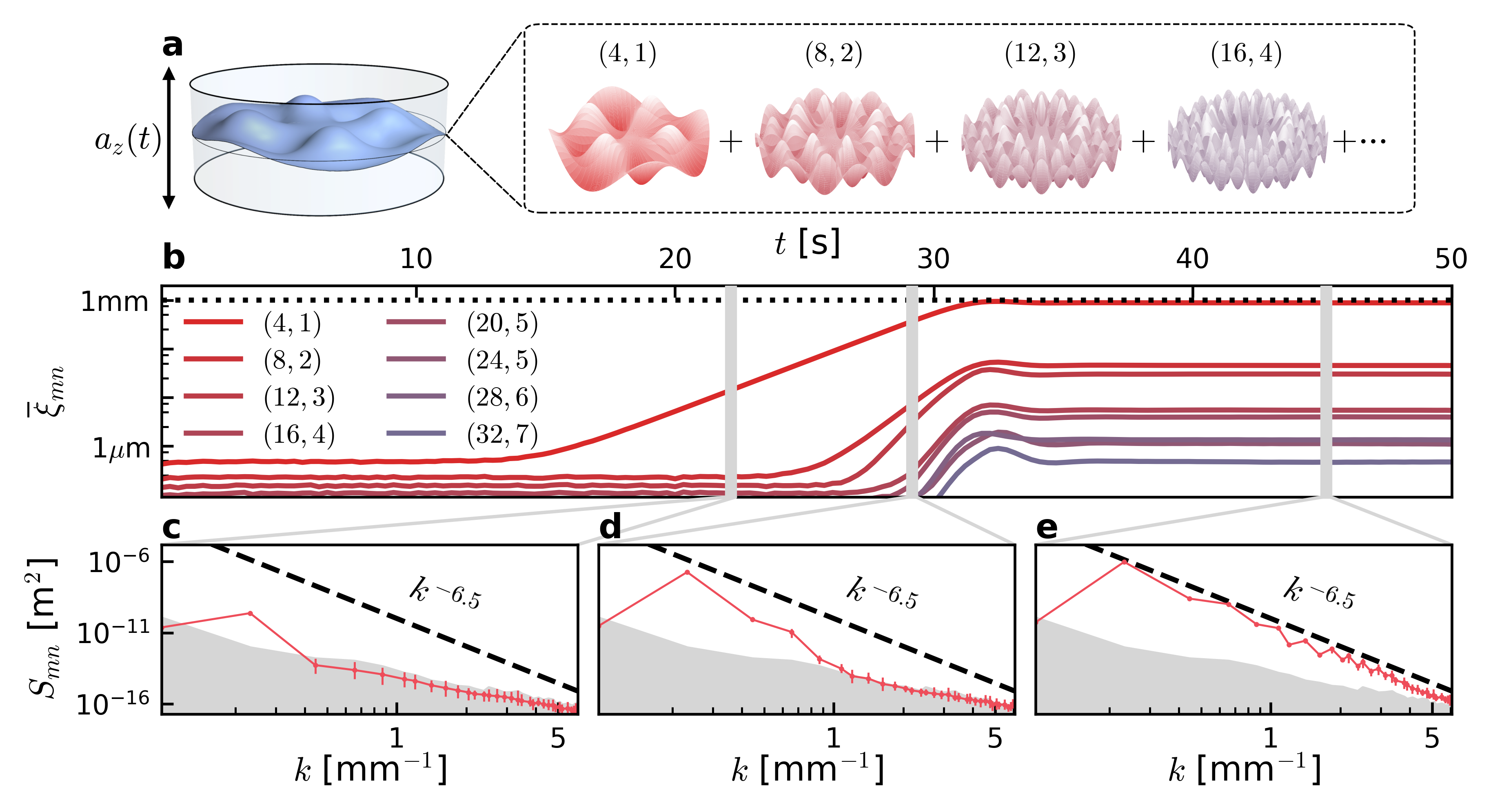}
\caption{\textbf{Observed interfacial wave dynamics.} In \textbf{a} (left), we show the measured interface between the biphasic solution of water, ethanol and potassium carbonate, filling the cylindrical cell of radius $r_0=\SI{40}{\milli\meter}$ and height $2h_0=\SI{30}{\milli\meter}$, periodically driven by a vertical acceleration $a_z(t)$. The resting interface (thin circular contour) is located halfway through the cell. In \textbf{a} (right), we depict the decomposition of the interface height $\xi$ in terms of spatial profiles $\Psi_{mn}$ for wave modes $(m,n)$ with the lowest wavenumbers $k_{mn}$ observed in the experiment. In \textbf{b}, the time evolutions of particular $(m,n)$ mode-amplitude envelopes $\bar{\xi}_{mn}$ are presented. The predicted saturation value of $\overline{\xi}_{4,1}$ is marked with a dotted line (Methods). Panels \textbf{c-e} show the power spectral density (PSD), $S_{mn}$, at three stages of the evolution. In \textbf{c}, at $t\sim \SI{22}{\second}$, only the primary $(4,1)$ mode is excited above the measurement noise level (grey-shaded region). In \textbf{d}, at $t\sim\SI{29}{\second}$, this one mode then interacts with a few others sourcing their growth. In \textbf{e}, at $t\sim\SI{45}{\second}$, the few excited modes interact with many more, and a quasi-stationary regime is established, as the spectrum develops a power-law behaviour $k^{-\alpha}$ indicative of a direct energy cascade. Across \textbf{c-e}, error bars indicate the uncertainty across experimental repetitions, and dashed lines with an $\alpha=6.5$ trend are presented for comparison.}
\label{fig:one}
\end{figure*}

For the controlled and repeatable creation of an out-of-equilibrium steady state, we start with the resting interface between two layers of immiscible liquids that fill a cylindrical cell (Fig.~\ref{fig:one}\textbf{a}). A bespoke spring-mass platform vertically drives the cell with acceleration $a_z(t)$, causing waves to appear on the interface with their spatial profiles determined by the cylindrical geometry. Our core observable is the two-fluid interface, whose three-dimensional profile is accessible over time using synthetic Schlieren imaging~\cite{Wildeman2018Real-timeBackdrop}.  Along the azimuthal direction $\theta$, the interfacial waves are periodic and can be identified by their number $m$ of crests or troughs. For each azimuthal wave $m$, the confining rigid outer wall discretises the available transverse radial profiles $R_{mn}(r)$, labelled by an integer $n$ (Methods). We show in Fig.~\ref{fig:one}\textbf{a} the spatial profile (real part), $\Psi_{mn} (r,\theta)=R_{mn}(r) \exp(im \theta)$, of the first few resolved interfacial wave modes $(m,n)$ with the lowest wavenumbers $k_{mn}$, or longest wavelengths. 

We model the time-evolving interface height $\xi(t,r,\theta)$ as a linear combination of the spatial profiles of wave modes $\Psi_{mn}(r,\theta)$ with associated time-dependent amplitudes $\xi_{mn}(t)$, that is, $\xi = \sum_{m,n} \xi_{mn}\Psi_{mn}$. From the measured time-evolving interface, we obtain rapidly oscillating amplitudes $\xi_{mn}(t)$, whose behaviour over large timescales is determined by their time-averaged envelopes $\bar{\xi}_{mn}$ (Methods), shown in Fig.~\ref{fig:one}\textbf{b} for a set of observed interfacial wave modes. 

Initially, the mode amplitudes fall below our measurement resolution threshold, and their envelopes appear approximately flat over time (see early times in Fig.~\ref{fig:one}\textbf{b}). At $t=\SI{0}{\second}$, the driver starts to oscillate the cell with acceleration $a_z(t)=a_0\sin(2\pi f_d t)$ of frequency $f_d=\SI{6.21}{\hertz}$ and amplitude $a_0=\SI{2.53(4)}{\meter\per\second^2}$, creating standing waves at the interface with frequency at integer multiples of $f_0=f_d/2$ and with exponentially increasing amplitude over time, a process known as parametric resonance~\cite{kumar1994parametric}. With our choice of parameters, the driver sources a wave at scales comparable to the diameter of the cell, with mode numbers $(m,n)=(4,1)$ and angular frequency $\omega_0=2\pi f_0$. We denote this wave by $(m,n)_{\omega}=(4,1)_{\omega_0}$ and refer to it as the \emph{primary} mode from here on. As it undergoes parametric resonance, the growth of the primary $(4,1)_{\omega_0}$ becomes evident in Fig.~\ref{fig:one}\textbf{c}, where the envelope $\bar{\xi}_{4,1}$ grows exponentially above the measurement resolution threshold from~$\sim\SI{13}{\second}$. 

At early times, the power spectral density (PSD), $S_{mn}=|\xi_{mn}|^2$, in Fig.~\ref{fig:one}\textbf{c} shows strong occupation of the primary wavenumber $k_{4,1}=\SI{0.23}{\per\milli\meter}$. By continuously driving the cell, the amplitude of the large-lengthscale primary (low wavenumber $k$) continues to grow until it becomes a source for shorter-wavelength (higher $k$) interfacial waves. These wave-mixing processes act to transfer energy from the primary to a few \emph{secondary} modes, thereby sourcing their exponential growth seen in Fig.~\ref{fig:one}\textbf{b} from around~$\SI{25}{\second}$, and increasing the occupation of higher wavenumbers in the PSD (Fig.~\ref{fig:one}\textbf{d}). These secondaries, in turn, source many more modes, stabilising at stationary amplitudes, while energy cascades to increasingly higher wavenumbers (Fig.~\ref{fig:one}\textbf{e}). Ultimately, constant energy flux is established across length-scales as the energy transferred to each mode balances both energy converted to heat (via viscous dissipation) and energy transported away to higher $k$ (via direct cascade)~\cite{nazarenko2011wave,FalconReview22}. Once a steady state is established, we observe a spectral cascade emerge in Fig.~\ref{fig:one}\textbf{e} with a characteristic power-law scaling $S_{mn}\sim k^{-\alpha}$~\cite{obukhov1941spectral}, where $\alpha=6.62(18)$.

The theory of interacting (or nonlinear) interfacial-wave dynamics underlying the formation of spectral cascades has been extensively investigated in fluid dynamics~\cite{miles1984nonlinear,miles1977Hamilton,nazarenko2011wave,nazarenko2016review}. In particular, a description through an effective Lagrangian field theory~\cite{miles1986NonlinearStratified,Barroso_2023,Nazarenko2015} offers a powerful interpretation of the interacting evolution in terms of scattering processes between interfacial wave modes. The effective Lagrangian $L$ for our system is given as an expansion in powers of the small wave amplitudes $\xi_{mn}$, that is, $L=\sum_{N\geq 2} L^{(N)}$, where $L^{(N)}$ corresponds to the $N$-th order term in the expansion (Methods). 

The quadratic term $L^{(2)}$ determines the independent (or linear) evolution of interfacial waves and reads
\begin{equation}
    L^{(2)}=\frac{1}{2}\sum_{a}\left[ c_a\left(|\dot\xi_a|^2-\omega^2_a|\xi_a|^2\right)+ a_z(t)\mathrm{At}|\xi_a|^2\right],    
\end{equation}
where $\omega_a$ is the dispersion frequency of a mode $a=(m_a,n_a)$ with wavenumber $k_a$, $\mathrm{At}$ is the Atwood number and $c_a=[k_a\tanh(k_ah_0)]^{-1}$. The non-negligible viscous dissipation of our fluids can be incorporated into this description~\cite{Barroso_2023}, yielding a damped linear equation of motion for individual modes independently evolving~\cite{kumar1994parametric,kovacic2018mathieu}. The resulting equation gives us predictive power on the modes that undergo parametric resonance and their exponential growth rate. For the primary mode $(4,1)_{\omega_0}$, the observed growth rate between $22$ and $29$ seconds in Fig.~\ref{fig:one}\textbf{b}, $\lambda^{\text{exp}}_{4,1}=\SI{0.467(8)}{\per\second}$, closely agrees with the prediction $\lambda^{\text{pred}}_{4,1}=\SI{0.46(1)}{\per\second}$. 

Higher-order Lagrangian terms determine the possible wave-mixing interactions. For instance, consider the cubic order term, 
\begin{equation}
    L^{(3)}=\frac{1}{2}\sum_{a,b,c}f_{abc} \xi_{a}\dot\xi_{b}\dot\xi_{c}.
\end{equation}
The strength of interaction between any three modes $a,b,c=(m_a,n_a),(m_b,n_b),(m_c,n_c)$ is determined by the magnitude of the coupling coefficient $f_{abc}$ defined as an integral over spatial eigenfunctions $\Psi_{mn}(r,\theta)$ (Methods). We find this coefficient to be proportional to the Kronecker delta of azimuthal numbers $f_{abc}\propto \delta_{m_a+m_b+m_c,0}$. This may be interpreted as a conservation law: the azimuthal numbers of all modes involved in the interaction must sum to zero. Coupling coefficients of higher-order terms are also proportional to Kronecker deltas of $m$, and so this conservation law generalises to all orders $L^{(N)}$. In similar fashion, one can determine a general conservation of oscillation frequency~\cite{nazarenko2011wave}.

\begin{figure*}[t!]
\centering
\includegraphics[width=\textwidth]{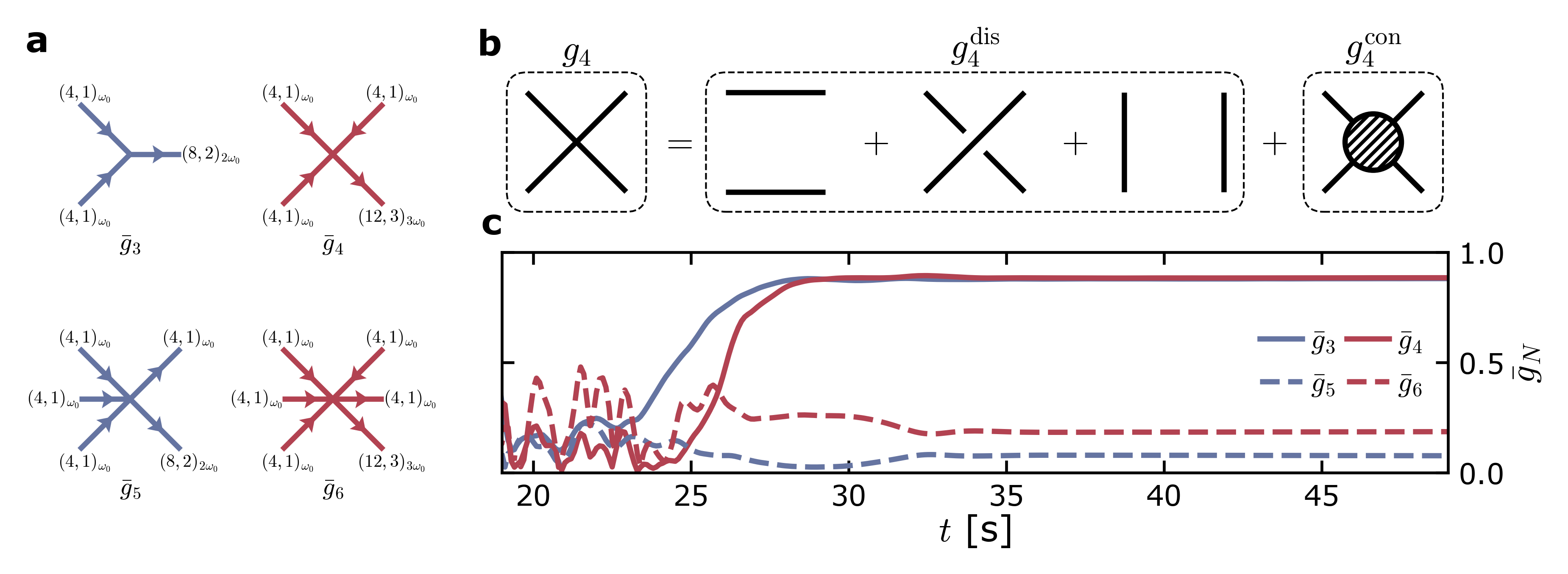}
\caption{\textbf{Scattering diagrams for wave mixing.} In \textbf{a} are presented the Feynman diagrams of the interactions between the primary $(4,1)_{\omega_0}$ and the secondaries $(8,2)_{2\omega_0}$ and $(12,3)_{3\omega_0}$ up to sixth order. The interaction order is given by the number of legs connected to the central vertex. Incoming (outgoing) waves have arrows pointing towards (away from) the vertex. In \textbf{b}, we depict the decomposition of the fourth-order correlation $g_4$ in terms of disconnected correlations $g_4^{\text{dis}}$, which are broken down into all possible lower-order two-point correlations, and connected correlations $g_4^{\text{con}}$, whose diagram displays a hashed, circular region representing all possible processes connecting the four outer legs. Panel \textbf{c} shows the time evolution of the experimental correlation measures $\bar{g}_3$ and $\bar{g}_4$, whose line colours match their corresponding scattering diagrams. We also show subleading higher-order processes related to the same secondaries, namely $\bar{g}_5$ and $\bar{g}_6$. }
\label{fig:few}
\end{figure*}

In the framework of field theory, one may interpret interactions as scattering processes adhering to vertex conservation rules. We consider positive (negative) azimuthal numbers to be incoming (outgoing) modes, with amplitudes related by complex conjugation $\xi_{-m,n}=\xi_{m,n}^*$.
For example, the process $(4,1)_{\omega_0}+(4,1)_{\omega_0}\to(8,2)_{2\omega_0}$ corresponds to the $\xi_{4,1}^{2}\xi_{-8,2}$ term of $L^{(3)}$. Feynman diagrams serve as an intuitive visual representation for these interactions~\cite{rosenhaus2023feynman}, and are shown in Fig.~\ref{fig:few}\textbf{a}, for the dominant processes involving the primary mode from the third to sixth orders. The secondary modes taking part in these interactions can be predicted by numerically evaluating the Lagrangian coupling coefficients, which show that the two modes most significantly excited by the primary are the $(8,2)_{2\omega_0}$ and $(12,3)_{3\omega_0}$. This is consistent with the first secondaries we observe growing above the noise level after $t\sim\SI{25}{\second}$ in Fig.~\ref{fig:one}\textbf{b}. To verify that the growth of these secondaries is indeed due to the specific interactions predicted from the Lagrangian, we employ the statistical machinery of correlation functions. 

To stress the importance of the conserved quantities, azimuthal number $m$ and frequency $\omega$, they are used to label interface waves $\xi_{m,\omega}(t,r)$, leaving the radial index $n$ implicit. Here, for a positive azimuthal number $m$, a positive (negative) frequency wave is one travelling counterclockwise (clockwise) azimuthally. We use the quantity $\xi_{m,\omega}$ to introduce, along the lines of~\cite{schweigler2017experimental}, a measure of interactions between collective degrees of freedom $\xi_{i}\equiv\xi_{m_i,\omega_i}$, namely the correlation functions $g_N(\xi_{1}\dots \xi_{N}) = \langle \xi_{1}\dots \xi_{N}\rangle$.
Here, the average $\langle\cdot\rangle$ is computed over spatial dimensions and experimental repetitions, rendering $g_N$ a function of time.
In general, this measure can be decomposed as $g_N = g^{\text{dis}}_N+g^{\text{con}}_N$ where, at each order $N$, only the connected part $g^{\text{con}}_N$, or statistical cumulant, contains new information, while the disconnected part $g^{\text{dis}}_N$ is fully characterised by lower-order correlations. A graphical depiction of the fourth-order decomposition is displayed in Fig.~\ref{fig:few}\textbf{b}.

In Fig.~\ref{fig:few}\textbf{c}, we show the normalised form of the connected correlation function $\bar{g}_N(t)$ (Methods) with which we can track a particular interaction over time. We choose to examine the dominant interactions depicted in Fig.~\ref{fig:few}\textbf{a}: $\bar{g}_3(\xi_{4,\omega_0}^2\xi_{-8,2\omega_0})$, $\bar{g}_4(\xi_{4,\omega_0}^3\xi_{-12,3\omega_0})$, $\bar{g}_5(\xi_{4,\omega_0}^3 \xi_{-4,\omega_0} \xi_{-8,2\omega_0})$, and $\bar{g}_6(\xi_{4,\omega_0}^4 \xi_{-4,\omega_0} \xi_{-12,3\omega_0})$. 
We observe both the cubic, $\bar{g}_3$, and quartic, $\bar{g}_4$, terms become significant when the primary mode is sufficiently large to source the amplification of the secondaries beyond the measurement threshold. This occurs around $\SI{25}{\second}$, when $\bar{g}_3$ and $\bar{g}_4$ grow significantly. Both correlations increase rapidly until they become maximally correlated at $\SI{28}{\second}$, before any of the mode amplitudes have saturated and when secondary modes trigger further interactions (see Fig.~\ref{fig:one}\textbf{c}). 

In contrast to the third and fourth-order interactions, we observe the fifth $\bar{g}_5$ and sixth $\bar{g}_6$ (and higher orders not presented) to be negligible throughout the nonlinear evolution, thus the majority of information regarding scattering processes is captured by $\bar{g}_3$ or $\bar{g}_4$. This finding justifies a truncation of the nonlinear theory at quartic order, limiting necessary analysis to the the potentials $L^{(3)}$ and $L^{(4)}$. As a result, the large coherent amplitude of the mode $(4,1)_{\omega_0}$ sources the equations of motion of the secondaries with quadratic and cubic powers of $\xi_{4,\omega_0}$.

We further verify these statements by investigating processes occurring at late times, when all available modes are excited and interact maximally. At this stage, interfacial waves reach their peak amplitude and oscillate at multiples of $\omega_0$. Numerous possible interactions contribute to the growth of each mode, and the sheer number of these contributions makes it impractical to track each one individually through exact diagrams. For this reason, we focus on correlations $G_N$ describing scattering processes with $N$ frequency legs, with each leg representing the ensemble of all waves oscillating at a frequency $\omega_i$. Similar to the approach used in Fig. \ref{fig:few}, we compute a normalised connected correlation measure, $\bar{G}_N(\omega_1,\dots,\omega_N)$ (Methods), averaged over the time interval from $\SI{43}{\second}$ to $\SI{48}{\second}$ within the highly nonlinear regime. 

\begin{figure}[t!]
\centering
\includegraphics[scale=1]{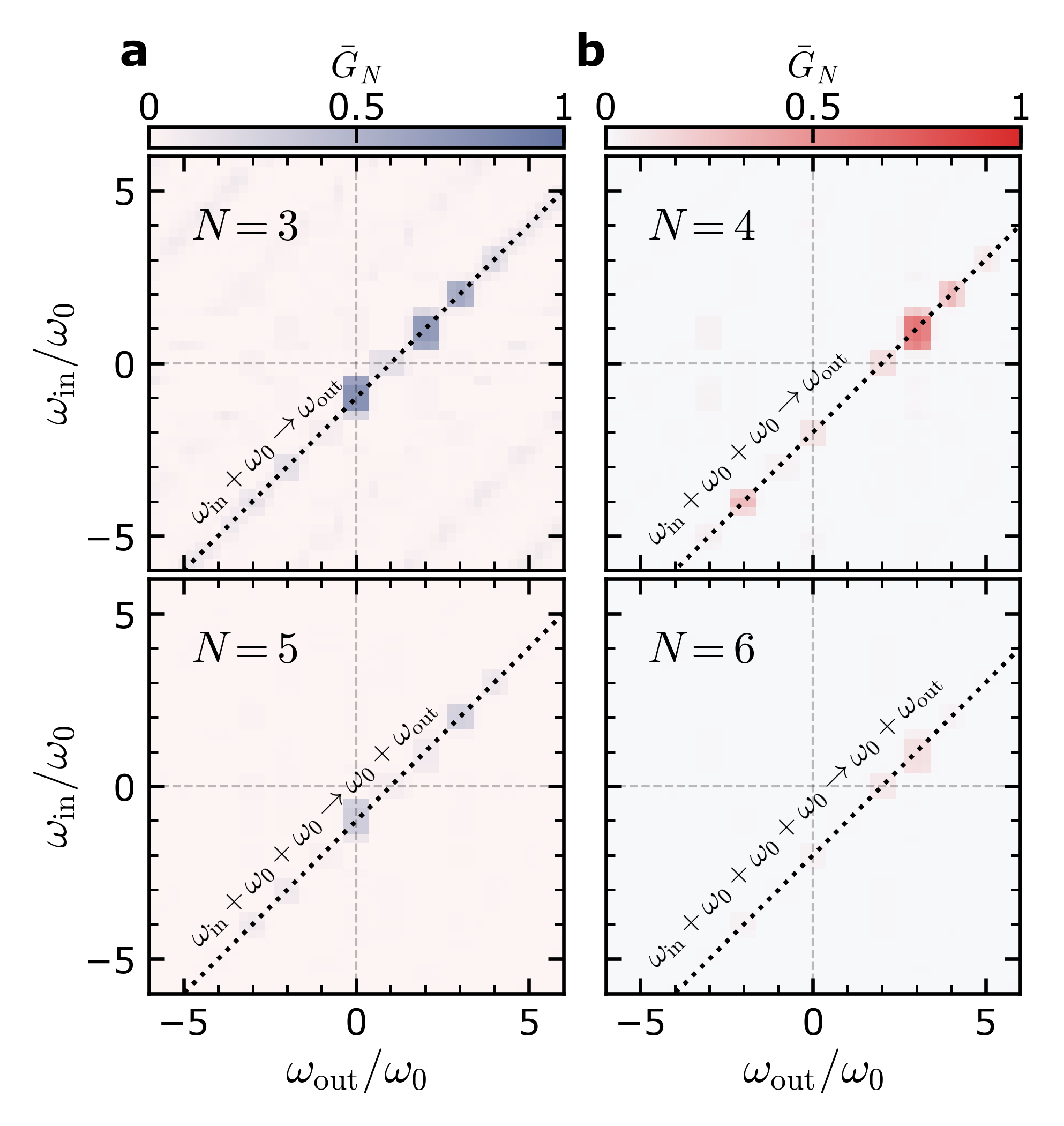}
\caption{\textbf{Interaction hierarchy.} The four panels represent different orders of the normalised correlation functions $\bar{G}_N$ at late time, encompassing all possible processes of the forms displayed in Fig.~\ref{fig:few}\textbf{a}. Two frequencies are left free, $\omega_{\textrm{in}}$ and $\omega_{\textrm{out}}$. The remaining terms of the interaction are fixed at $\omega_0$. The expected directions of frequency conservation are depicted by dotted lines. All measures are shown from zero (light) to one (dark). In the top row, the processes $\omega_0+\omega_0\to 2\omega_0$ in \textbf{a} and $\omega_0+\omega_0+\omega_0\to 3\omega_0$ in \textbf{b} appear as dark spots. According to the conservation of the azimuthal number, order $N=3$ contains the contribution of all cubic interactions conserving frequencies, e.g. $(4,1)_{\omega_0}+(8,2)_{2\omega_0}\to(12,3)_{3\omega_0}$ or $(20,n)_{\omega_0}+(16,n')_{2\omega_0}\to(36,n'')_{3\omega_0}$. The contribution of $\bar{G}_5$ and $\bar{G}_6$ in the bottom row, hence of the fifth and sixth-order interactions, appear significantly smaller than the lower orders.}
\label{fig:many}
\end{figure}

We present the $\bar{G}_N$ measure in Fig.~\ref{fig:many} for orders $N=3,4,5,6$, where one ingoing frequency, $\omega_{\textrm{in}}$, and one outgoing frequency, $\omega_{\textrm{out}}$, are free to vary, while all other remaining terms are fixed to $\omega_0$. We observe high correlation at isolated peaks (dark spots) indicating an interaction process between modes with those frequencies. In every order, these interaction points are located only along the line of frequency conservation, at which the sum of incoming frequencies equals that of outgoing ones. We also see the correlations of $\bar{G}_3$ and $\bar{G}_4$ (upper panels of Fig.~\ref{fig:many}) to be significantly larger than correlations at higher orders. Indeed, the $\bar{G}_5$ and $\bar{G}_6$ display very weak correlation (lower panels of Fig.~\ref{fig:many}), which suggests the terms $L^{(5)}$ and $L^{(6)}$ remain negligible in the far-from-equilibrium steady state. This result is consistent with weak wave turbulence assumptions~\cite{FalconReview22}, and further justifies truncating the system's Lagrangian at fourth order. 

\begin{figure}[t!]
\centering
\includegraphics[scale=1]{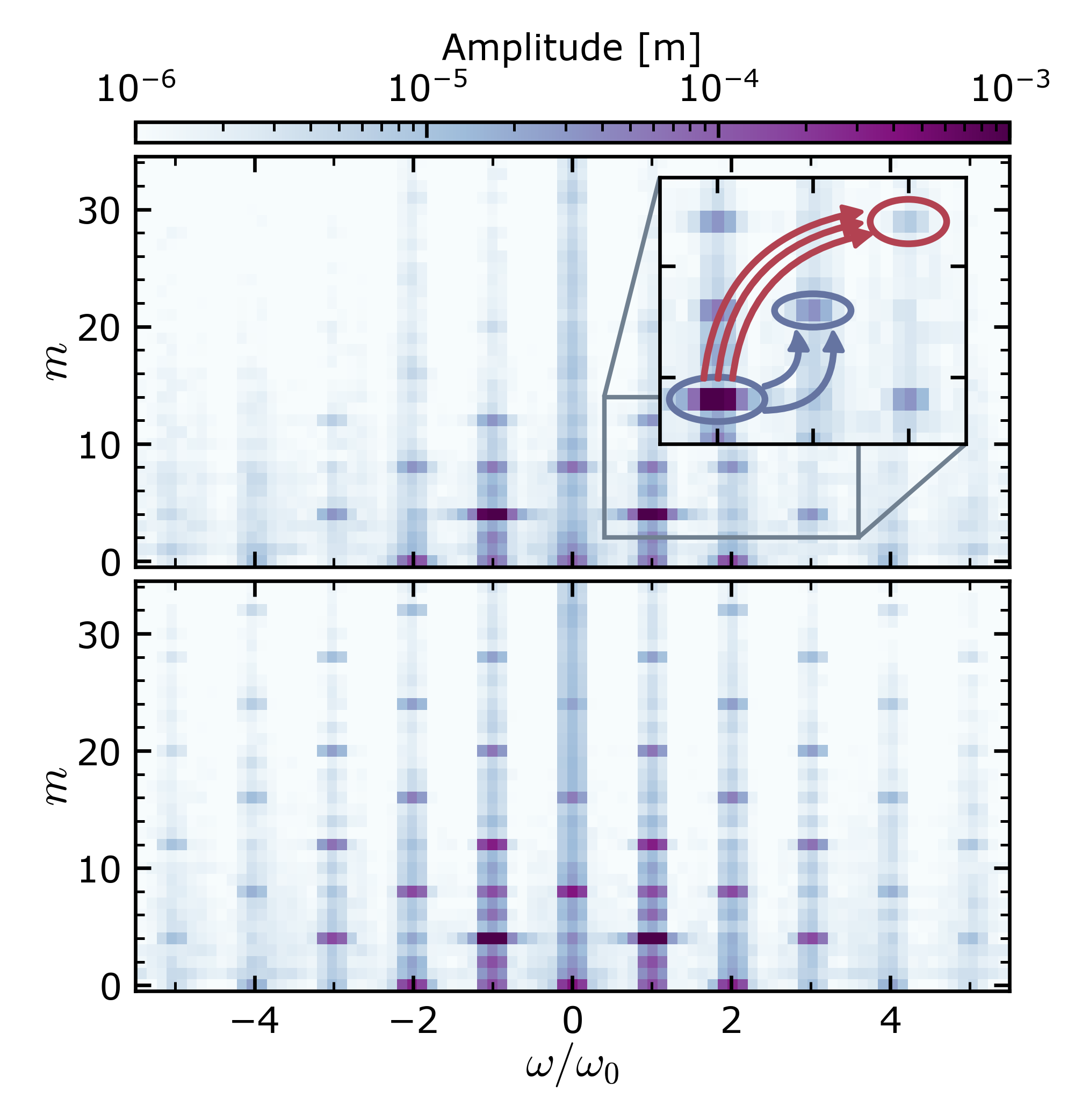}
\caption{\textbf{Cascade tree.} Top and bottom show the magnitude of the amplitudes $|\xi_{m,\omega}|$ averaged over $r$ and a time interval of one second around early (top, $t\sim\SI{30}{\second}$) and late (bottom, $t\sim\SI{45}{\second}$) times of the evolution. \textbf{Top}, the inset depicts the processes corresponding to the cubic (blue) and quartic (red) mixing between modes as in Fig.~\ref{fig:few}\textbf{a}. \textbf{Bottom}, the amplitude map shows a larger number of the $(m,\omega)$ excited modes in this regime with respect to the top panel.}
\label{fig:cascade}
\end{figure}

Having demonstrated all significant interactions to be of third or fourth order, conserving both frequency and azimuthal number, we no longer need to consider every possible interaction to determine which modes become excited at late times. Instead, we have a predictive rule: such interactions only trigger modes at either even or odd integer multiples of both the frequency and azimuthal number of the primary. We observe this rule to hold true from the alternating pattern of interface waves $\xi_{m,\omega}$ appearing in Fig.~\ref{fig:cascade}.
The top panel presents an example from the early stages of evolution, where only a limited number of secondary modes are excited as a result of the interaction of the primary mode with itself.
The bottom panel represents a later time point, during which the cluster disperses across broader frequencies and azimuthal numbers in a cascade-tree manner~\cite{l2010discrete} until a steady state is established.

This cascading evolution, also observed in the PSD (Fig.~\ref{fig:one}\textbf{(c-e)}), is comprehensively explained by the interaction picture we have presented.
The energy transport from the injection scale to smaller lengthscales continues over time, travelling along channels of cubic and quartic interactions with conserved frequency and azimuthal number. Ultimately, the growth of the primary is arrested through nonlinear self-interaction. An approximate single-mode dynamics for the primary wave may be obtained from a reduced quartic Lagrangian~\cite{miles1984nonlinear}; yielding a prediction for the steady-state saturation amplitude of the primary (Methods),
\begin{equation} \label{eqn:SatAmp}
\overline{\xi}_{a}^{(\mathrm{Sat})} = \mathcal{C} \left[ \omega_0^2 - \omega_{a}^2 + \sqrt{ \left(\frac{a_0\mathrm{At}}{2 c_a} \right)^2 - \left( 2 \omega_0 \gamma_{a}\right)^2} \right]^{1/2}
\end{equation}

where $\gamma_a$ is the damping rate and $\mathcal{C}$ is a parameter inversely proportional to the relevant coupling constants. For $a=(4,1)$, Eq.~\ref{eqn:SatAmp} predicts $\overline{\xi}_{4,1}^{(\mathrm{Sat})} =\SI{1.01\pm0.06}{\milli\meter}$ which is marked as a dotted line against the value measured, $\SI{0.98\pm0.04}{\milli \meter}$, in Fig~\ref{fig:one}\textbf{c}. 
Once the primary reaches this amplitude, all other modes rapidly stop growing as the energy transported to them balances that transported away to larger $k$, and that lost to viscous dissipation. The timescale until this steady-state is reached from an initial state with amplitude $\overline{\xi}_{4,1}^{(\mathrm{I})}$ is $\Delta t = \lambda_{4,1}^{-1} \log \left(\overline{\xi}_{4,1}^{(\mathrm{Sat})}/\overline{\xi}_{4,1}^{(\mathrm{I})}\right)$.

During the late time steady-state, the PSD follows a power-law trend $S(k) \sim k^{-\alpha}$, the characteristic signature of a direct energy cascade~\cite{obukhov1941spectral}. While this phenomenon is common to isotropic wave turbulence, it lacks detailed empirical treatment under strong forcing and finite-size conditions, both present within our experiment. 
The highly discretised set of available spatial profiles resulting from the bounding volume of the cell, together with a forcing quantified as strong by the wave steepness~\cite{FalconReview22}, $k_{4,1}\xi_{\mathrm{rms}}= 0.33 >0.1$, leads to depletion of wave resonances, causing secondary modes to deviate from their dispersion frequencies~\cite{l2010discrete,snouck2009turbulent}. Since dispersion-preserving processes are an key assumption of weak wave turbulence, the theory is not expected to apply to a system such as ours, marking a surprising result to observe a direct cascade prevailing despite such conditions.

We have demonstrated that our system has the control and precision required to extract the intricate structure of out-of-equilibrium phenomena, including the formation of spectral cascades and the determination of their characteristic power laws. 
These features are ubiquitous in a variety of far-from-equilibrium systems. They play a key role in the theory of reheating~\cite{kofman1997towards}, which is required to understand how the early Universe transitions from cold to thermal; connecting the end of cosmic inflation to the Hot Big Bang model of standard cosmology. As we have previously shown in~\cite{barroso2022primary}, our hydrodynamical system provides a simulator for the early, resonant phase of cosmological preheating. Here, we confirm that the analogy holds in the far-from-equilibrium regime and show that the wave dynamics is well-characterised by a truncated effective Lagrangian. As a consequence, we can establish a translation key, mapping the parameters of the hydrodynamical model to those of the cosmological target system. We can exploit this correspondence for physical simulations of prominent early-Universe scenarios using our system. 

In cosmological reheating, the hypothetical scalar field driving inflation undergoes a phase of rapid oscillations. Similarly to the mechanical driver in our system, the scalar field transfers its energy to other matter fields through parametric resonance, thus producing particles that eventually thermalise. For generic inflationary models~\cite{Martin:2013tda,kaiser2016nonminimal}, the interacting dynamics of these matter fields is described by a Lagrangian closely resembling that of interfacial waves in our experiment (Methods). Numerical simulations of such reheating dynamics find very similar behaviour \cite{LozanovAmin2017oscillons,nguyen2019nonlinear,van2020time,ShafCopelMahbMishraBas}  
to what we have observed here, forming cascades between modes of differing wavenumber $k$. To date, however, such cosmological studies have only been able to explore the early onset of (pre)thermalisation, and the full mechanism by which a thermal equilibrium emerges from such far-from-equilibrium dynamics remains beyond the reach of current state-of-the-art cosmological lattice simulations~\cite{amin2015nonperturbative,nguyen2019nonlinear,van2020time,giblin2019preheating,Figueroa:2020rrl}. Such intricate regimes are readily accessible in our hydrodynamical simulator, offering a complementary route to explore thermalisation mechanisms. Indeed, one can map tuneable parameters of our experimental system (driving frequency, amplitude, and hydrodynamical properties) to fundamental parameters of the inflationary model (inflationary potential, Lagrangian couplings), so that the experiment becomes a powerful emulator of the physics of cosmological reheating. For instance, $L^{(3)}$ can be further suppressed by reducing the difference in fluid densities, allowing for a direct correspondence with a quartic Lagrangian conventional for theories of preheating.

Our work presents a comprehensive experimental analysis on the formation of a direct energy cascade. From the appearance of scaling behavior within the context of strong forcing, dissipation, and finite-size effects, we learn that this phenomenon is more robust against theoretically challenging conditions than previously appreciated~\cite{FalconReview22}. A truncated Lagrangian is used to model the formation process enabling the calculation of dominant interactions, which serve as channels for energy to be transported throughout the system, and the timescale, through a reduced dynamics model. We verify the accuracy of both by tracking the evolution of individual wave modes, via geometric decomposition, and using field-theoretical correlation measures to examine interactions between them. Our methodology presents exciting new prospects for the study of universal nonequilibrium behaviours, complementing recent successes from cold atom experiments~\cite{prufer2018observation,Erne_2018,gazo2023universal,martirosyan2024universal} with the advantage of precise time resolution over individual experimental runs. 

Detailed understanding of the formation of far-from-equilibrium steady-states, as presented, will ultimately enable us to control nonlinear systems. By injecting specific modes and subsequently triggering specific interactions, we can guide or prevent a journey into the complex landscape of far-from-equilibrium physics.


\paragraph{\textbf{Acknowledgements}}
We thank Sebastian Erne, Christoph Eigen, Thomas Gasenzer, Sergey Nazarenko, Kostas Kokkotas and members of the Gravity Laboratory for comments and suggestions. We thank Paul Saffin, Ed Copeland, Swagat Mishra, Oli Gould, Mustafa Amin and Tom Giblin for discussions on the cosmological implications of this work.  
We acknowledge the support provided by the Leverhulme Research Leadership Award (RL-2019-020),
the Royal Society University Research Fellowship
(UF120112,RF/ERE/210198, RGF/EA/180286, RGF/EA/181015), and partial
support by the Science and Technology Facilities Council (Theory Consolidated Grant ST/P000703/1), the Science and Technology Facilities Council on Quantum Simulators for Fundamental Physics (ST/T006900/1) as part
of the UKRI Quantum Technologies for Fundamental
Physics programme. Portions of this work were conducted in MIT's Center for Theoretical Physics and supported in part by the U.~S.~Department of Energy under Contract No.~DE-SC0012567.

\paragraph{\textbf{Data Availability}}
The data supporting the findings of this study are available from the corresponding author upon reasonable request.

\paragraph{\textbf{Authors contributions}}
All authors contributed substantially to the work. Overall conceptualisation by A.A. and S.W., and experimental design by V.B. and S.W.. S.G. and S.S. performed the experiment with V.B. providing experimental guidance, S.G., S.S. and V.B. performed the data analysis and made the numerical calculations. A.A., V.B., D.K. and S.W. provided theoretical guidance. All authors contributed to interpreting the data and writing the manuscript. S.G. and S.S. contributed equally.

\paragraph{\textbf{Competing interests}} The authors declare no competing interests.

\paragraph{\textbf{Correspondence and requests for materials}} should be addressed to Silke Weinfurtner.

\bibliography{Biblio}

\begin{thebibliography}{67}%
\makeatletter
\providecommand \@ifxundefined [1]{%
 \@ifx{#1\undefined}
}%
\providecommand \@ifnum [1]{%
 \ifnum #1\expandafter \@firstoftwo
 \else \expandafter \@secondoftwo
 \fi
}%
\providecommand \@ifx [1]{%
 \ifx #1\expandafter \@firstoftwo
 \else \expandafter \@secondoftwo
 \fi
}%
\providecommand \natexlab [1]{#1}%
\providecommand \enquote  [1]{``#1''}%
\providecommand \bibnamefont  [1]{#1}%
\providecommand \bibfnamefont [1]{#1}%
\providecommand \citenamefont [1]{#1}%
\providecommand \href@noop [0]{\@secondoftwo}%
\providecommand \href [0]{\begingroup \@sanitize@url \@href}%
\providecommand \@href[1]{\@@startlink{#1}\@@href}%
\providecommand \@@href[1]{\endgroup#1\@@endlink}%
\providecommand \@sanitize@url [0]{\catcode `\\12\catcode `\$12\catcode `\&12\catcode `\#12\catcode `\^12\catcode `\_12\catcode `\%12\relax}%
\providecommand \@@startlink[1]{}%
\providecommand \@@endlink[0]{}%
\providecommand \url  [0]{\begingroup\@sanitize@url \@url }%
\providecommand \@url [1]{\endgroup\@href {#1}{\urlprefix }}%
\providecommand \urlprefix  [0]{URL }%
\providecommand \Eprint [0]{\href }%
\providecommand \doibase [0]{https://doi.org/}%
\providecommand \selectlanguage [0]{\@gobble}%
\providecommand \bibinfo  [0]{\@secondoftwo}%
\providecommand \bibfield  [0]{\@secondoftwo}%
\providecommand \translation [1]{[#1]}%
\providecommand \BibitemOpen [0]{}%
\providecommand \bibitemStop [0]{}%
\providecommand \bibitemNoStop [0]{.\EOS\space}%
\providecommand \EOS [0]{\spacefactor3000\relax}%
\providecommand \BibitemShut  [1]{\csname bibitem#1\endcsname}%
\let\auto@bib@innerbib\@empty
\bibitem [{\citenamefont {Lovejoy}(2018)}]{lovejoy2018spectra}%
  \BibitemOpen
  \bibfield  {author} {\bibinfo {author} {\bibfnamefont {S.}~\bibnamefont {Lovejoy}},\ }\bibfield  {title} {\bibinfo {title} {Spectra, intermittency, and extremes of weather, macroweather and climate},\ }\href@noop {} {\bibfield  {journal} {\bibinfo  {journal} {Scientific reports}\ }\textbf {\bibinfo {volume} {8}},\ \bibinfo {pages} {12697} (\bibinfo {year} {2018})}\BibitemShut {NoStop}%
\bibitem [{\citenamefont {Ghashghaie}\ \emph {et~al.}(1996)\citenamefont {Ghashghaie}, \citenamefont {Breymann}, \citenamefont {Peinke}, \citenamefont {Talkner},\ and\ \citenamefont {Dodge}}]{ghashghaie1996turbulent}%
  \BibitemOpen
  \bibfield  {author} {\bibinfo {author} {\bibfnamefont {S.}~\bibnamefont {Ghashghaie}}, \bibinfo {author} {\bibfnamefont {W.}~\bibnamefont {Breymann}}, \bibinfo {author} {\bibfnamefont {J.}~\bibnamefont {Peinke}}, \bibinfo {author} {\bibfnamefont {P.}~\bibnamefont {Talkner}},\ and\ \bibinfo {author} {\bibfnamefont {Y.}~\bibnamefont {Dodge}},\ }\bibfield  {title} {\bibinfo {title} {Turbulent cascades in foreign exchange markets},\ }\href@noop {} {\bibfield  {journal} {\bibinfo  {journal} {Nature}\ }\textbf {\bibinfo {volume} {381}},\ \bibinfo {pages} {767} (\bibinfo {year} {1996})}\BibitemShut {NoStop}%
\bibitem [{\citenamefont {Sheremet}\ \emph {et~al.}(2019)\citenamefont {Sheremet}, \citenamefont {Qin}, \citenamefont {Kennedy}, \citenamefont {Zhou},\ and\ \citenamefont {Maurer}}]{sheremet2019wave}%
  \BibitemOpen
  \bibfield  {author} {\bibinfo {author} {\bibfnamefont {A.}~\bibnamefont {Sheremet}}, \bibinfo {author} {\bibfnamefont {Y.}~\bibnamefont {Qin}}, \bibinfo {author} {\bibfnamefont {J.~P.}\ \bibnamefont {Kennedy}}, \bibinfo {author} {\bibfnamefont {Y.}~\bibnamefont {Zhou}},\ and\ \bibinfo {author} {\bibfnamefont {A.~P.}\ \bibnamefont {Maurer}},\ }\bibfield  {title} {\bibinfo {title} {Wave turbulence and energy cascade in the hippocampus},\ }\href@noop {} {\bibfield  {journal} {\bibinfo  {journal} {Frontiers in systems neuroscience}\ }\textbf {\bibinfo {volume} {12}},\ \bibinfo {pages} {62} (\bibinfo {year} {2019})}\BibitemShut {NoStop}%
\bibitem [{\citenamefont {Zakharov}\ \emph {et~al.}(2012)\citenamefont {Zakharov}, \citenamefont {L'vov},\ and\ \citenamefont {Falkovich}}]{zakharov2012kolmogorov}%
  \BibitemOpen
  \bibfield  {author} {\bibinfo {author} {\bibfnamefont {V.~E.}\ \bibnamefont {Zakharov}}, \bibinfo {author} {\bibfnamefont {V.~S.}\ \bibnamefont {L'vov}},\ and\ \bibinfo {author} {\bibfnamefont {G.}~\bibnamefont {Falkovich}},\ }\href@noop {} {\emph {\bibinfo {title} {Kolmogorov spectra of turbulence I: Wave turbulence}}}\ (\bibinfo  {publisher} {Springer Science \& Business Media},\ \bibinfo {year} {2012})\BibitemShut {NoStop}%
\bibitem [{\citenamefont {{Falcon}}\ and\ \citenamefont {{Mordant}}(2022)}]{FalconReview22}%
  \BibitemOpen
  \bibfield  {author} {\bibinfo {author} {\bibfnamefont {E.}~\bibnamefont {{Falcon}}}\ and\ \bibinfo {author} {\bibfnamefont {N.}~\bibnamefont {{Mordant}}},\ }\bibfield  {title} {\bibinfo {title} {{Experiments in Surface Gravity-Capillary Wave Turbulence}},\ }\href {https://doi.org/10.1146/annurev-fluid-021021-102043} {\bibfield  {journal} {\bibinfo  {journal} {Annual Review of Fluid Mechanics}\ }\textbf {\bibinfo {volume} {54}},\ \bibinfo {pages} {1} (\bibinfo {year} {2022})}\BibitemShut {NoStop}%
\bibitem [{\citenamefont {{Miles}}(1986)}]{miles1986NonlinearStratified}%
  \BibitemOpen
  \bibfield  {author} {\bibinfo {author} {\bibfnamefont {J.~W.}\ \bibnamefont {{Miles}}},\ }\bibfield  {title} {\bibinfo {title} {{Weakly nonlinear waves in a stratified fluid: a variational formulation}},\ }\href {https://doi.org/10.1017/s0022112086001830} {\bibfield  {journal} {\bibinfo  {journal} {Journal of Fluid Mechanics}\ }\textbf {\bibinfo {volume} {172}},\ \bibinfo {pages} {499} (\bibinfo {year} {1986})}\BibitemShut {NoStop}%
\bibitem [{\citenamefont {Barroso}\ \emph {et~al.}(2023)\citenamefont {Barroso}, \citenamefont {Bunney},\ and\ \citenamefont {Weinfurtner}}]{Barroso_2023}%
  \BibitemOpen
  \bibfield  {author} {\bibinfo {author} {\bibfnamefont {V.~S.}\ \bibnamefont {Barroso}}, \bibinfo {author} {\bibfnamefont {C.~R.~D.}\ \bibnamefont {Bunney}},\ and\ \bibinfo {author} {\bibfnamefont {S.}~\bibnamefont {Weinfurtner}},\ }\bibfield  {title} {\bibinfo {title} {Non-linear effective field theory simulators in two-fluid interfaces},\ }\href {https://doi.org/10.1088/1742-6596/2531/1/012003} {\bibfield  {journal} {\bibinfo  {journal} {Journal of Physics: Conference Series}\ }\textbf {\bibinfo {volume} {2531}},\ \bibinfo {pages} {012003} (\bibinfo {year} {2023})}\BibitemShut {NoStop}%
\bibitem [{\citenamefont {Nazarenko}(2011)}]{nazarenko2011wave}%
  \BibitemOpen
  \bibfield  {author} {\bibinfo {author} {\bibfnamefont {S.}~\bibnamefont {Nazarenko}},\ }\href@noop {} {\emph {\bibinfo {title} {Wave turbulence}}},\ Vol.\ \bibinfo {volume} {825}\ (\bibinfo  {publisher} {Springer},\ \bibinfo {year} {2011})\BibitemShut {NoStop}%
\bibitem [{\citenamefont {Kofman}\ \emph {et~al.}(1997)\citenamefont {Kofman}, \citenamefont {Linde},\ and\ \citenamefont {Starobinsky}}]{kofman1997towards}%
  \BibitemOpen
  \bibfield  {author} {\bibinfo {author} {\bibfnamefont {L.}~\bibnamefont {Kofman}}, \bibinfo {author} {\bibfnamefont {A.}~\bibnamefont {Linde}},\ and\ \bibinfo {author} {\bibfnamefont {A.~A.}\ \bibnamefont {Starobinsky}},\ }\bibfield  {title} {\bibinfo {title} {Towards the theory of reheating after inflation},\ }\href@noop {} {\bibfield  {journal} {\bibinfo  {journal} {Physical Review D}\ }\textbf {\bibinfo {volume} {56}},\ \bibinfo {pages} {3258} (\bibinfo {year} {1997})}\BibitemShut {NoStop}%
\bibitem [{\citenamefont {Amin}\ \emph {et~al.}(2014)\citenamefont {Amin}, \citenamefont {Hertzberg}, \citenamefont {Kaiser},\ and\ \citenamefont {Karouby}}]{amin2015nonperturbative}%
  \BibitemOpen
  \bibfield  {author} {\bibinfo {author} {\bibfnamefont {M.~A.}\ \bibnamefont {Amin}}, \bibinfo {author} {\bibfnamefont {M.~P.}\ \bibnamefont {Hertzberg}}, \bibinfo {author} {\bibfnamefont {D.~I.}\ \bibnamefont {Kaiser}},\ and\ \bibinfo {author} {\bibfnamefont {J.}~\bibnamefont {Karouby}},\ }\bibfield  {title} {\bibinfo {title} {{Nonperturbative Dynamics Of Reheating After Inflation: A Review}},\ }\href {https://doi.org/10.1142/s0218271815300037} {\bibfield  {journal} {\bibinfo  {journal} {Int. J. Mod. Phys. D}\ }\textbf {\bibinfo {volume} {24}},\ \bibinfo {pages} {1530003} (\bibinfo {year} {2014})}\BibitemShut {NoStop}%
\bibitem [{\citenamefont {{Nazarenko}}\ and\ \citenamefont {{Lukaschuk}}(2016)}]{nazarenko2016review}%
  \BibitemOpen
  \bibfield  {author} {\bibinfo {author} {\bibfnamefont {S.}~\bibnamefont {{Nazarenko}}}\ and\ \bibinfo {author} {\bibfnamefont {S.}~\bibnamefont {{Lukaschuk}}},\ }\bibfield  {title} {\bibinfo {title} {{Wave Turbulence on Water Surface}},\ }\href {https://doi.org/10.1146/annurev-conmatphys-071715-102737} {\bibfield  {journal} {\bibinfo  {journal} {Annual Review of Condensed Matter Physics}\ }\textbf {\bibinfo {volume} {7}},\ \bibinfo {pages} {61} (\bibinfo {year} {2016})}\BibitemShut {NoStop}%
\bibitem [{\citenamefont {Yamada}\ \emph {et~al.}(2008)\citenamefont {Yamada}, \citenamefont {Itoh}, \citenamefont {Maruta}, \citenamefont {Kasuya}, \citenamefont {Nagashima}, \citenamefont {Shinohara}, \citenamefont {Terasaka}, \citenamefont {Yagi}, \citenamefont {Inagaki}, \citenamefont {Kawai} \emph {et~al.}}]{yamada2008anatomy}%
  \BibitemOpen
  \bibfield  {author} {\bibinfo {author} {\bibfnamefont {T.}~\bibnamefont {Yamada}}, \bibinfo {author} {\bibfnamefont {S.-I.}\ \bibnamefont {Itoh}}, \bibinfo {author} {\bibfnamefont {T.}~\bibnamefont {Maruta}}, \bibinfo {author} {\bibfnamefont {N.}~\bibnamefont {Kasuya}}, \bibinfo {author} {\bibfnamefont {Y.}~\bibnamefont {Nagashima}}, \bibinfo {author} {\bibfnamefont {S.}~\bibnamefont {Shinohara}}, \bibinfo {author} {\bibfnamefont {K.}~\bibnamefont {Terasaka}}, \bibinfo {author} {\bibfnamefont {M.}~\bibnamefont {Yagi}}, \bibinfo {author} {\bibfnamefont {S.}~\bibnamefont {Inagaki}}, \bibinfo {author} {\bibfnamefont {Y.}~\bibnamefont {Kawai}}, \emph {et~al.},\ }\bibfield  {title} {\bibinfo {title} {Anatomy of plasma turbulence},\ }\href@noop {} {\bibfield  {journal} {\bibinfo  {journal} {Nature physics}\ }\textbf {\bibinfo {volume} {4}},\ \bibinfo {pages} {721} (\bibinfo {year} {2008})}\BibitemShut {NoStop}%
\bibitem [{\citenamefont {Wyngaard}(1992)}]{wyngaard1992atmospheric}%
  \BibitemOpen
  \bibfield  {author} {\bibinfo {author} {\bibfnamefont {J.~C.}\ \bibnamefont {Wyngaard}},\ }\bibfield  {title} {\bibinfo {title} {Atmospheric turbulence},\ }\href@noop {} {\bibfield  {journal} {\bibinfo  {journal} {Annual Review of Fluid Mechanics}\ }\textbf {\bibinfo {volume} {24}},\ \bibinfo {pages} {205} (\bibinfo {year} {1992})}\BibitemShut {NoStop}%
\bibitem [{\citenamefont {Abdurakhimov}\ \emph {et~al.}(2010)\citenamefont {Abdurakhimov}, \citenamefont {Brazhnikov}, \citenamefont {Remizov},\ and\ \citenamefont {Levchenko}}]{abdurakhimov2010observation}%
  \BibitemOpen
  \bibfield  {author} {\bibinfo {author} {\bibfnamefont {L.~V.}\ \bibnamefont {Abdurakhimov}}, \bibinfo {author} {\bibfnamefont {M.~Y.}\ \bibnamefont {Brazhnikov}}, \bibinfo {author} {\bibfnamefont {I.}~\bibnamefont {Remizov}},\ and\ \bibinfo {author} {\bibfnamefont {A.~A.}\ \bibnamefont {Levchenko}},\ }\bibfield  {title} {\bibinfo {title} {Observation of wave energy accumulation in the turbulent spectrum of capillary waves on the he-ii surface under harmonic pumping},\ }\href@noop {} {\bibfield  {journal} {\bibinfo  {journal} {JETP letters}\ }\textbf {\bibinfo {volume} {91}},\ \bibinfo {pages} {271} (\bibinfo {year} {2010})}\BibitemShut {NoStop}%
\bibitem [{\citenamefont {Abdurakhimov}\ \emph {et~al.}(2011)\citenamefont {Abdurakhimov}, \citenamefont {Brazhnikov}, \citenamefont {Remizov},\ and\ \citenamefont {Levchenko}}]{abdurakhimov2011classical}%
  \BibitemOpen
  \bibfield  {author} {\bibinfo {author} {\bibfnamefont {L.}~\bibnamefont {Abdurakhimov}}, \bibinfo {author} {\bibfnamefont {M.~Y.}\ \bibnamefont {Brazhnikov}}, \bibinfo {author} {\bibfnamefont {I.}~\bibnamefont {Remizov}},\ and\ \bibinfo {author} {\bibfnamefont {A.}~\bibnamefont {Levchenko}},\ }\bibfield  {title} {\bibinfo {title} {Classical capillary turbulence on the surface of quantum liquid he-ii},\ }\href@noop {} {\bibfield  {journal} {\bibinfo  {journal} {Low Temperature Physics}\ }\textbf {\bibinfo {volume} {37}},\ \bibinfo {pages} {403} (\bibinfo {year} {2011})}\BibitemShut {NoStop}%
\bibitem [{\citenamefont {Abdurakhimov}\ \emph {et~al.}(2012)\citenamefont {Abdurakhimov}, \citenamefont {Brazhnikov}, \citenamefont {Levchenko}, \citenamefont {Remizov},\ and\ \citenamefont {Filatov}}]{abdurakhimov2012turbulent}%
  \BibitemOpen
  \bibfield  {author} {\bibinfo {author} {\bibfnamefont {L.~V.}\ \bibnamefont {Abdurakhimov}}, \bibinfo {author} {\bibfnamefont {M.~Y.}\ \bibnamefont {Brazhnikov}}, \bibinfo {author} {\bibfnamefont {A.~A.}\ \bibnamefont {Levchenko}}, \bibinfo {author} {\bibfnamefont {I.}~\bibnamefont {Remizov}},\ and\ \bibinfo {author} {\bibfnamefont {S.~V.}\ \bibnamefont {Filatov}},\ }\bibfield  {title} {\bibinfo {title} {Turbulent capillary cascade near the edge of the inertial range on the surface of a quantum liquid},\ }\href@noop {} {\bibfield  {journal} {\bibinfo  {journal} {JETP letters}\ }\textbf {\bibinfo {volume} {95}},\ \bibinfo {pages} {670} (\bibinfo {year} {2012})}\BibitemShut {NoStop}%
\bibitem [{\citenamefont {Navon}\ \emph {et~al.}(2016)\citenamefont {Navon}, \citenamefont {Gaunt}, \citenamefont {Smith},\ and\ \citenamefont {Hadzibabic}}]{navon2016emergence}%
  \BibitemOpen
  \bibfield  {author} {\bibinfo {author} {\bibfnamefont {N.}~\bibnamefont {Navon}}, \bibinfo {author} {\bibfnamefont {A.~L.}\ \bibnamefont {Gaunt}}, \bibinfo {author} {\bibfnamefont {R.~P.}\ \bibnamefont {Smith}},\ and\ \bibinfo {author} {\bibfnamefont {Z.}~\bibnamefont {Hadzibabic}},\ }\bibfield  {title} {\bibinfo {title} {Emergence of a turbulent cascade in a quantum gas},\ }\href@noop {} {\bibfield  {journal} {\bibinfo  {journal} {Nature}\ }\textbf {\bibinfo {volume} {539}},\ \bibinfo {pages} {72} (\bibinfo {year} {2016})}\BibitemShut {NoStop}%
\bibitem [{\citenamefont {Navon}\ \emph {et~al.}(2019)\citenamefont {Navon}, \citenamefont {Eigen}, \citenamefont {Zhang}, \citenamefont {Lopes}, \citenamefont {Gaunt}, \citenamefont {Fujimoto}, \citenamefont {Tsubota}, \citenamefont {Smith},\ and\ \citenamefont {Hadzibabic}}]{navon2019synthetic}%
  \BibitemOpen
  \bibfield  {author} {\bibinfo {author} {\bibfnamefont {N.}~\bibnamefont {Navon}}, \bibinfo {author} {\bibfnamefont {C.}~\bibnamefont {Eigen}}, \bibinfo {author} {\bibfnamefont {J.}~\bibnamefont {Zhang}}, \bibinfo {author} {\bibfnamefont {R.}~\bibnamefont {Lopes}}, \bibinfo {author} {\bibfnamefont {A.~L.}\ \bibnamefont {Gaunt}}, \bibinfo {author} {\bibfnamefont {K.}~\bibnamefont {Fujimoto}}, \bibinfo {author} {\bibfnamefont {M.}~\bibnamefont {Tsubota}}, \bibinfo {author} {\bibfnamefont {R.~P.}\ \bibnamefont {Smith}},\ and\ \bibinfo {author} {\bibfnamefont {Z.}~\bibnamefont {Hadzibabic}},\ }\bibfield  {title} {\bibinfo {title} {Synthetic dissipation and cascade fluxes in a turbulent quantum gas},\ }\href@noop {} {\bibfield  {journal} {\bibinfo  {journal} {Science}\ }\textbf {\bibinfo {volume} {366}},\ \bibinfo {pages} {382} (\bibinfo {year} {2019})}\BibitemShut {NoStop}%
\bibitem [{\citenamefont {Zhang}\ \emph {et~al.}(2021)\citenamefont {Zhang}, \citenamefont {Eigen}, \citenamefont {Zheng}, \citenamefont {Glidden}, \citenamefont {Hilker}, \citenamefont {Garratt}, \citenamefont {Lopes}, \citenamefont {Cooper}, \citenamefont {Hadzibabic},\ and\ \citenamefont {Navon}}]{zhang2021many}%
  \BibitemOpen
  \bibfield  {author} {\bibinfo {author} {\bibfnamefont {J.}~\bibnamefont {Zhang}}, \bibinfo {author} {\bibfnamefont {C.}~\bibnamefont {Eigen}}, \bibinfo {author} {\bibfnamefont {W.}~\bibnamefont {Zheng}}, \bibinfo {author} {\bibfnamefont {J.~A.}\ \bibnamefont {Glidden}}, \bibinfo {author} {\bibfnamefont {T.~A.}\ \bibnamefont {Hilker}}, \bibinfo {author} {\bibfnamefont {S.~J.}\ \bibnamefont {Garratt}}, \bibinfo {author} {\bibfnamefont {R.}~\bibnamefont {Lopes}}, \bibinfo {author} {\bibfnamefont {N.~R.}\ \bibnamefont {Cooper}}, \bibinfo {author} {\bibfnamefont {Z.}~\bibnamefont {Hadzibabic}},\ and\ \bibinfo {author} {\bibfnamefont {N.}~\bibnamefont {Navon}},\ }\bibfield  {title} {\bibinfo {title} {Many-body decay of the gapped lowest excitation of a bose-einstein condensate},\ }\href@noop {} {\bibfield  {journal} {\bibinfo  {journal} {Physical Review Letters}\ }\textbf {\bibinfo {volume} {126}},\ \bibinfo {pages} {060402} (\bibinfo {year} {2021})}\BibitemShut {NoStop}%
\bibitem [{\citenamefont {Glidden}\ \emph {et~al.}(2021)\citenamefont {Glidden}, \citenamefont {Eigen}, \citenamefont {Dogra}, \citenamefont {Hilker}, \citenamefont {Smith},\ and\ \citenamefont {Hadzibabic}}]{glidden2021bidirectional}%
  \BibitemOpen
  \bibfield  {author} {\bibinfo {author} {\bibfnamefont {J.~A.}\ \bibnamefont {Glidden}}, \bibinfo {author} {\bibfnamefont {C.}~\bibnamefont {Eigen}}, \bibinfo {author} {\bibfnamefont {L.~H.}\ \bibnamefont {Dogra}}, \bibinfo {author} {\bibfnamefont {T.~A.}\ \bibnamefont {Hilker}}, \bibinfo {author} {\bibfnamefont {R.~P.}\ \bibnamefont {Smith}},\ and\ \bibinfo {author} {\bibfnamefont {Z.}~\bibnamefont {Hadzibabic}},\ }\bibfield  {title} {\bibinfo {title} {Bidirectional dynamic scaling in an isolated bose gas far from equilibrium},\ }\href@noop {} {\bibfield  {journal} {\bibinfo  {journal} {Nature Physics}\ }\textbf {\bibinfo {volume} {17}},\ \bibinfo {pages} {457} (\bibinfo {year} {2021})}\BibitemShut {NoStop}%
\bibitem [{\citenamefont {Ga{\l}ka}\ \emph {et~al.}(2022)\citenamefont {Ga{\l}ka}, \citenamefont {Christodoulou}, \citenamefont {Gazo}, \citenamefont {Karailiev}, \citenamefont {Dogra}, \citenamefont {Schmitt},\ and\ \citenamefont {Hadzibabic}}]{Ga_ka_2022}%
  \BibitemOpen
  \bibfield  {author} {\bibinfo {author} {\bibfnamefont {M.}~\bibnamefont {Ga{\l}ka}}, \bibinfo {author} {\bibfnamefont {P.}~\bibnamefont {Christodoulou}}, \bibinfo {author} {\bibfnamefont {M.}~\bibnamefont {Gazo}}, \bibinfo {author} {\bibfnamefont {A.}~\bibnamefont {Karailiev}}, \bibinfo {author} {\bibfnamefont {N.}~\bibnamefont {Dogra}}, \bibinfo {author} {\bibfnamefont {J.}~\bibnamefont {Schmitt}},\ and\ \bibinfo {author} {\bibfnamefont {Z.}~\bibnamefont {Hadzibabic}},\ }\bibfield  {title} {\bibinfo {title} {Emergence of isotropy and dynamic scaling in 2d wave turbulence in a homogeneous bose gas},\ }\bibfield  {journal} {\bibinfo  {journal} {Physical Review Letters}\ }\textbf {\bibinfo {volume} {129}},\ \href {https://doi.org/10.1103/physrevlett.129.190402} {10.1103/physrevlett.129.190402} (\bibinfo {year} {2022})\BibitemShut {NoStop}%
\bibitem [{\citenamefont {Dogra}\ \emph {et~al.}(2023)\citenamefont {Dogra}, \citenamefont {Martirosyan}, \citenamefont {Hilker}, \citenamefont {Glidden}, \citenamefont {Etrych}, \citenamefont {Cao}, \citenamefont {Eigen}, \citenamefont {Smith},\ and\ \citenamefont {Hadzibabic}}]{dogra2023universal}%
  \BibitemOpen
  \bibfield  {author} {\bibinfo {author} {\bibfnamefont {L.~H.}\ \bibnamefont {Dogra}}, \bibinfo {author} {\bibfnamefont {G.}~\bibnamefont {Martirosyan}}, \bibinfo {author} {\bibfnamefont {T.~A.}\ \bibnamefont {Hilker}}, \bibinfo {author} {\bibfnamefont {J.~A.}\ \bibnamefont {Glidden}}, \bibinfo {author} {\bibfnamefont {J.}~\bibnamefont {Etrych}}, \bibinfo {author} {\bibfnamefont {A.}~\bibnamefont {Cao}}, \bibinfo {author} {\bibfnamefont {C.}~\bibnamefont {Eigen}}, \bibinfo {author} {\bibfnamefont {R.~P.}\ \bibnamefont {Smith}},\ and\ \bibinfo {author} {\bibfnamefont {Z.}~\bibnamefont {Hadzibabic}},\ }\bibfield  {title} {\bibinfo {title} {Universal equation of state for wave turbulence in a quantum gas},\ }\href@noop {} {\bibfield  {journal} {\bibinfo  {journal} {Nature}\ }\textbf {\bibinfo {volume} {620}},\ \bibinfo {pages} {521} (\bibinfo {year} {2023})}\BibitemShut {NoStop}%
\bibitem [{\citenamefont {Koniakhin}\ \emph {et~al.}(2020)\citenamefont {Koniakhin}, \citenamefont {Bleu}, \citenamefont {Malpuech},\ and\ \citenamefont {Solnyshkov}}]{KONIAKHIN2020109574}%
  \BibitemOpen
  \bibfield  {author} {\bibinfo {author} {\bibfnamefont {S.}~\bibnamefont {Koniakhin}}, \bibinfo {author} {\bibfnamefont {O.}~\bibnamefont {Bleu}}, \bibinfo {author} {\bibfnamefont {G.}~\bibnamefont {Malpuech}},\ and\ \bibinfo {author} {\bibfnamefont {D.}~\bibnamefont {Solnyshkov}},\ }\bibfield  {title} {\bibinfo {title} {2d quantum turbulence in a polariton quantum fluid},\ }\href {https://doi.org/https://doi.org/10.1016/j.chaos.2019.109574} {\bibfield  {journal} {\bibinfo  {journal} {Chaos, Solitons \& Fractals}\ }\textbf {\bibinfo {volume} {132}},\ \bibinfo {pages} {109574} (\bibinfo {year} {2020})}\BibitemShut {NoStop}%
\bibitem [{\citenamefont {Baker-Rasooli}\ \emph {et~al.}(2023)\citenamefont {Baker-Rasooli}, \citenamefont {Liu}, \citenamefont {Aladjidi}, \citenamefont {Bramati},\ and\ \citenamefont {Glorieux}}]{baker2023turbulent}%
  \BibitemOpen
  \bibfield  {author} {\bibinfo {author} {\bibfnamefont {M.}~\bibnamefont {Baker-Rasooli}}, \bibinfo {author} {\bibfnamefont {W.}~\bibnamefont {Liu}}, \bibinfo {author} {\bibfnamefont {T.}~\bibnamefont {Aladjidi}}, \bibinfo {author} {\bibfnamefont {A.}~\bibnamefont {Bramati}},\ and\ \bibinfo {author} {\bibfnamefont {Q.}~\bibnamefont {Glorieux}},\ }\bibfield  {title} {\bibinfo {title} {Turbulent dynamics in a two-dimensional paraxial fluid of light},\ }\href@noop {} {\bibfield  {journal} {\bibinfo  {journal} {Physical Review A}\ }\textbf {\bibinfo {volume} {108}},\ \bibinfo {pages} {063512} (\bibinfo {year} {2023})}\BibitemShut {NoStop}%
\bibitem [{\citenamefont {Wildeman}(2018)}]{Wildeman2018Real-timeBackdrop}%
  \BibitemOpen
  \bibfield  {author} {\bibinfo {author} {\bibfnamefont {S.}~\bibnamefont {Wildeman}},\ }\bibfield  {title} {\bibinfo {title} {{Real-time quantitative Schlieren imaging by fast Fourier demodulation of a checkered backdrop}},\ }\href {https://doi.org/10.1007/s00348-018-2553-9} {\bibfield  {journal} {\bibinfo  {journal} {Experiments in Fluids}\ }\textbf {\bibinfo {volume} {59}},\ \bibinfo {pages} {97} (\bibinfo {year} {2018})}\BibitemShut {NoStop}%
\bibitem [{\citenamefont {Kumar}\ and\ \citenamefont {Tuckerman}(1994)}]{kumar1994parametric}%
  \BibitemOpen
  \bibfield  {author} {\bibinfo {author} {\bibfnamefont {K.}~\bibnamefont {Kumar}}\ and\ \bibinfo {author} {\bibfnamefont {L.~S.}\ \bibnamefont {Tuckerman}},\ }\bibfield  {title} {\bibinfo {title} {Parametric instability of the interface between two fluids},\ }\href@noop {} {\bibfield  {journal} {\bibinfo  {journal} {Journal of Fluid Mechanics}\ }\textbf {\bibinfo {volume} {279}},\ \bibinfo {pages} {49} (\bibinfo {year} {1994})}\BibitemShut {NoStop}%
\bibitem [{\citenamefont {Obukhov}(1941)}]{obukhov1941spectral}%
  \BibitemOpen
  \bibfield  {author} {\bibinfo {author} {\bibfnamefont {A.}~\bibnamefont {Obukhov}},\ }\bibfield  {title} {\bibinfo {title} {Spectral energy distribution in a turbulent flow},\ }\href@noop {} {\bibfield  {journal} {\bibinfo  {journal} {Izv. Akad. Nauk. SSSR. Ser. Geogr. i. Geofiz}\ }\textbf {\bibinfo {volume} {5}},\ \bibinfo {pages} {453} (\bibinfo {year} {1941})}\BibitemShut {NoStop}%
\bibitem [{\citenamefont {Miles}(1984)}]{miles1984nonlinear}%
  \BibitemOpen
  \bibfield  {author} {\bibinfo {author} {\bibfnamefont {J.~W.}\ \bibnamefont {Miles}},\ }\bibfield  {title} {\bibinfo {title} {Nonlinear faraday resonance},\ }\href@noop {} {\bibfield  {journal} {\bibinfo  {journal} {Journal of Fluid Mechanics}\ }\textbf {\bibinfo {volume} {146}},\ \bibinfo {pages} {285} (\bibinfo {year} {1984})}\BibitemShut {NoStop}%
\bibitem [{\citenamefont {{Miles}}(1977)}]{miles1977Hamilton}%
  \BibitemOpen
  \bibfield  {author} {\bibinfo {author} {\bibfnamefont {J.~W.}\ \bibnamefont {{Miles}}},\ }\bibfield  {title} {\bibinfo {title} {{On Hamilton's principle for surface waves}},\ }\href {https://doi.org/10.1017/s0022112077001104} {\bibfield  {journal} {\bibinfo  {journal} {Journal of Fluid Mechanics}\ }\textbf {\bibinfo {volume} {83}},\ \bibinfo {pages} {153} (\bibinfo {year} {1977})}\BibitemShut {NoStop}%
\bibitem [{\citenamefont {Gallet}\ \emph {et~al.}(2015)\citenamefont {Gallet}, \citenamefont {Nazarenko},\ and\ \citenamefont {Dubrulle}}]{Nazarenko2015}%
  \BibitemOpen
  \bibfield  {author} {\bibinfo {author} {\bibfnamefont {B.}~\bibnamefont {Gallet}}, \bibinfo {author} {\bibfnamefont {S.}~\bibnamefont {Nazarenko}},\ and\ \bibinfo {author} {\bibfnamefont {B.}~\bibnamefont {Dubrulle}},\ }\bibfield  {title} {\bibinfo {title} {Wave-turbulence description of interacting particles: Klein-gordon model with a mexican-hat potential},\ }\href {https://doi.org/10.1103/PhysRevE.92.012909} {\bibfield  {journal} {\bibinfo  {journal} {Phys. Rev. E}\ }\textbf {\bibinfo {volume} {92}},\ \bibinfo {pages} {012909} (\bibinfo {year} {2015})}\BibitemShut {NoStop}%
\bibitem [{\citenamefont {Kovacic}\ \emph {et~al.}(2018)\citenamefont {Kovacic}, \citenamefont {Rand},\ and\ \citenamefont {Mohamed~Sah}}]{kovacic2018mathieu}%
  \BibitemOpen
  \bibfield  {author} {\bibinfo {author} {\bibfnamefont {I.}~\bibnamefont {Kovacic}}, \bibinfo {author} {\bibfnamefont {R.}~\bibnamefont {Rand}},\ and\ \bibinfo {author} {\bibfnamefont {S.}~\bibnamefont {Mohamed~Sah}},\ }\bibfield  {title} {\bibinfo {title} {Mathieu's equation and its generalizations: overview of stability charts and their features},\ }\href@noop {} {\bibfield  {journal} {\bibinfo  {journal} {Applied Mechanics Reviews}\ }\textbf {\bibinfo {volume} {70}} (\bibinfo {year} {2018})}\BibitemShut {NoStop}%
\bibitem [{\citenamefont {Rosenhaus}\ and\ \citenamefont {Smolkin}(2023)}]{rosenhaus2023feynman}%
  \BibitemOpen
  \bibfield  {author} {\bibinfo {author} {\bibfnamefont {V.}~\bibnamefont {Rosenhaus}}\ and\ \bibinfo {author} {\bibfnamefont {M.}~\bibnamefont {Smolkin}},\ }\bibfield  {title} {\bibinfo {title} {Feynman rules for forced wave turbulence},\ }\href@noop {} {\bibfield  {journal} {\bibinfo  {journal} {Journal of High Energy Physics}\ }\textbf {\bibinfo {volume} {2023}},\ \bibinfo {pages} {1} (\bibinfo {year} {2023})}\BibitemShut {NoStop}%
\bibitem [{\citenamefont {Schweigler}\ \emph {et~al.}(2017)\citenamefont {Schweigler}, \citenamefont {Kasper}, \citenamefont {Erne}, \citenamefont {Mazets}, \citenamefont {Rauer}, \citenamefont {Cataldini}, \citenamefont {Langen}, \citenamefont {Gasenzer}, \citenamefont {Berges},\ and\ \citenamefont {Schmiedmayer}}]{schweigler2017experimental}%
  \BibitemOpen
  \bibfield  {author} {\bibinfo {author} {\bibfnamefont {T.}~\bibnamefont {Schweigler}}, \bibinfo {author} {\bibfnamefont {V.}~\bibnamefont {Kasper}}, \bibinfo {author} {\bibfnamefont {S.}~\bibnamefont {Erne}}, \bibinfo {author} {\bibfnamefont {I.}~\bibnamefont {Mazets}}, \bibinfo {author} {\bibfnamefont {B.}~\bibnamefont {Rauer}}, \bibinfo {author} {\bibfnamefont {F.}~\bibnamefont {Cataldini}}, \bibinfo {author} {\bibfnamefont {T.}~\bibnamefont {Langen}}, \bibinfo {author} {\bibfnamefont {T.}~\bibnamefont {Gasenzer}}, \bibinfo {author} {\bibfnamefont {J.}~\bibnamefont {Berges}},\ and\ \bibinfo {author} {\bibfnamefont {J.}~\bibnamefont {Schmiedmayer}},\ }\bibfield  {title} {\bibinfo {title} {Experimental characterization of a quantum many-body system via higher-order correlations},\ }\href@noop {} {\bibfield  {journal} {\bibinfo  {journal} {Nature}\ }\textbf {\bibinfo {volume} {545}},\ \bibinfo {pages} {323} (\bibinfo {year} {2017})}\BibitemShut {NoStop}%
\bibitem [{\citenamefont {L’vov}\ and\ \citenamefont {Nazarenko}(2010)}]{l2010discrete}%
  \BibitemOpen
  \bibfield  {author} {\bibinfo {author} {\bibfnamefont {V.}~\bibnamefont {L’vov}}\ and\ \bibinfo {author} {\bibfnamefont {S.}~\bibnamefont {Nazarenko}},\ }\bibfield  {title} {\bibinfo {title} {Discrete and mesoscopic regimes of finite-size wave turbulence},\ }\href@noop {} {\bibfield  {journal} {\bibinfo  {journal} {Physical Review E—Statistical, Nonlinear, and Soft Matter Physics}\ }\textbf {\bibinfo {volume} {82}},\ \bibinfo {pages} {056322} (\bibinfo {year} {2010})}\BibitemShut {NoStop}%
\bibitem [{\citenamefont {Snouck}\ \emph {et~al.}(2009)\citenamefont {Snouck}, \citenamefont {Westra},\ and\ \citenamefont {van~de Water}}]{snouck2009turbulent}%
  \BibitemOpen
  \bibfield  {author} {\bibinfo {author} {\bibfnamefont {D.}~\bibnamefont {Snouck}}, \bibinfo {author} {\bibfnamefont {M.-T.}\ \bibnamefont {Westra}},\ and\ \bibinfo {author} {\bibfnamefont {W.}~\bibnamefont {van~de Water}},\ }\bibfield  {title} {\bibinfo {title} {Turbulent parametric surface waves},\ }\href@noop {} {\bibfield  {journal} {\bibinfo  {journal} {Physics of Fluids}\ }\textbf {\bibinfo {volume} {21}} (\bibinfo {year} {2009})}\BibitemShut {NoStop}%
\bibitem [{\citenamefont {Barroso}\ \emph {et~al.}(2022)\citenamefont {Barroso}, \citenamefont {Geelmuyden}, \citenamefont {Fifer}, \citenamefont {Erne}, \citenamefont {Avgoustidis}, \citenamefont {Hill},\ and\ \citenamefont {Weinfurtner}}]{barroso2022primary}%
  \BibitemOpen
  \bibfield  {author} {\bibinfo {author} {\bibfnamefont {V.~S.}\ \bibnamefont {Barroso}}, \bibinfo {author} {\bibfnamefont {A.}~\bibnamefont {Geelmuyden}}, \bibinfo {author} {\bibfnamefont {Z.}~\bibnamefont {Fifer}}, \bibinfo {author} {\bibfnamefont {S.}~\bibnamefont {Erne}}, \bibinfo {author} {\bibfnamefont {A.}~\bibnamefont {Avgoustidis}}, \bibinfo {author} {\bibfnamefont {R.~J.~A.}\ \bibnamefont {Hill}},\ and\ \bibinfo {author} {\bibfnamefont {S.}~\bibnamefont {Weinfurtner}},\ }\bibfield  {title} {\bibinfo {title} {Primary thermalisation mechanism of early universe observed from faraday-wave scattering on liquid-liquid interfaces},\ }\href@noop {} {\bibfield  {journal} {\bibinfo  {journal} {arXiv:2207.02199}\ } (\bibinfo {year} {2022})}\BibitemShut {NoStop}%
\bibitem [{\citenamefont {Martin}\ \emph {et~al.}(2024)\citenamefont {Martin}, \citenamefont {Ringeval},\ and\ \citenamefont {Vennin}}]{Martin:2013tda}%
  \BibitemOpen
  \bibfield  {author} {\bibinfo {author} {\bibfnamefont {J.}~\bibnamefont {Martin}}, \bibinfo {author} {\bibfnamefont {C.}~\bibnamefont {Ringeval}},\ and\ \bibinfo {author} {\bibfnamefont {V.}~\bibnamefont {Vennin}},\ }\bibfield  {title} {\bibinfo {title} {{Encyclop\ae{}dia Inflationaris: Opiparous Edition}},\ }\href {https://doi.org/https://doi.org/10.1016/j.dark.2024.101653} {\bibfield  {journal} {\bibinfo  {journal} {Physics of the Dark Universe}\ }\textbf {\bibinfo {volume} {46}},\ \bibinfo {pages} {101653} (\bibinfo {year} {2024})}\BibitemShut {NoStop}%
\bibitem [{\citenamefont {Kaiser}(2016)}]{kaiser2016nonminimal}%
  \BibitemOpen
  \bibfield  {author} {\bibinfo {author} {\bibfnamefont {D.~I.}\ \bibnamefont {Kaiser}},\ }\bibfield  {title} {\bibinfo {title} {{Nonminimal Couplings in the Early Universe: Multifield Models of Inflation and the Latest Observations}},\ }\href {https://doi.org/10.1007/978-3-319-31299-6_2} {\bibfield  {journal} {\bibinfo  {journal} {Fundam. Theor. Phys.}\ }\textbf {\bibinfo {volume} {183}},\ \bibinfo {pages} {41} (\bibinfo {year} {2016})}\BibitemShut {NoStop}%
\bibitem [{\citenamefont {Lozanov}\ and\ \citenamefont {Amin}(2018)}]{LozanovAmin2017oscillons}%
  \BibitemOpen
  \bibfield  {author} {\bibinfo {author} {\bibfnamefont {K.~D.}\ \bibnamefont {Lozanov}}\ and\ \bibinfo {author} {\bibfnamefont {M.~A.}\ \bibnamefont {Amin}},\ }\bibfield  {title} {\bibinfo {title} {{Self-resonance after inflation: oscillons, transients and radiation domination}},\ }\href {https://doi.org/10.1103/PhysRevD.97.023533} {\bibfield  {journal} {\bibinfo  {journal} {Phys. Rev. D}\ }\textbf {\bibinfo {volume} {97}},\ \bibinfo {pages} {023533} (\bibinfo {year} {2018})},\ \Eprint {https://arxiv.org/abs/1710.06851} {arXiv:1710.06851 [astro-ph.CO]} \BibitemShut {NoStop}%
\bibitem [{\citenamefont {Nguyen}\ \emph {et~al.}(2019)\citenamefont {Nguyen}, \citenamefont {van~de Vis}, \citenamefont {Sfakianakis}, \citenamefont {Giblin},\ and\ \citenamefont {Kaiser}}]{nguyen2019nonlinear}%
  \BibitemOpen
  \bibfield  {author} {\bibinfo {author} {\bibfnamefont {R.}~\bibnamefont {Nguyen}}, \bibinfo {author} {\bibfnamefont {J.}~\bibnamefont {van~de Vis}}, \bibinfo {author} {\bibfnamefont {E.~I.}\ \bibnamefont {Sfakianakis}}, \bibinfo {author} {\bibfnamefont {J.~T.}\ \bibnamefont {Giblin}},\ and\ \bibinfo {author} {\bibfnamefont {D.~I.}\ \bibnamefont {Kaiser}},\ }\bibfield  {title} {\bibinfo {title} {{Nonlinear Dynamics of Preheating after Multifield Inflation with Nonminimal Couplings}},\ }\href {https://doi.org/10.1103/PhysRevLett.123.171301} {\bibfield  {journal} {\bibinfo  {journal} {Phys. Rev. Lett.}\ }\textbf {\bibinfo {volume} {123}},\ \bibinfo {pages} {171301} (\bibinfo {year} {2019})}\BibitemShut {NoStop}%
\bibitem [{\citenamefont {van~de Vis}\ \emph {et~al.}(2020)\citenamefont {van~de Vis}, \citenamefont {Nguyen}, \citenamefont {Sfakianakis}, \citenamefont {Giblin},\ and\ \citenamefont {Kaiser}}]{van2020time}%
  \BibitemOpen
  \bibfield  {author} {\bibinfo {author} {\bibfnamefont {J.}~\bibnamefont {van~de Vis}}, \bibinfo {author} {\bibfnamefont {R.}~\bibnamefont {Nguyen}}, \bibinfo {author} {\bibfnamefont {E.~I.}\ \bibnamefont {Sfakianakis}}, \bibinfo {author} {\bibfnamefont {J.~T.}\ \bibnamefont {Giblin}},\ and\ \bibinfo {author} {\bibfnamefont {D.~I.}\ \bibnamefont {Kaiser}},\ }\bibfield  {title} {\bibinfo {title} {{Time scales for nonlinear processes in preheating after multifield inflation with nonminimal couplings}},\ }\href {https://doi.org/10.1103/PhysRevD.102.043528} {\bibfield  {journal} {\bibinfo  {journal} {Phys. Rev. D}\ }\textbf {\bibinfo {volume} {102}},\ \bibinfo {pages} {043528} (\bibinfo {year} {2020})}\BibitemShut {NoStop}%
\bibitem [{\citenamefont {Shafi}\ \emph {et~al.}(2024)\citenamefont {Shafi}, \citenamefont {Copeland}, \citenamefont {Mahbub}, \citenamefont {Mishra},\ and\ \citenamefont {Basak}}]{ShafCopelMahbMishraBas}%
  \BibitemOpen
  \bibfield  {author} {\bibinfo {author} {\bibfnamefont {M.}~\bibnamefont {Shafi}}, \bibinfo {author} {\bibfnamefont {E.~J.}\ \bibnamefont {Copeland}}, \bibinfo {author} {\bibfnamefont {R.}~\bibnamefont {Mahbub}}, \bibinfo {author} {\bibfnamefont {S.~S.}\ \bibnamefont {Mishra}},\ and\ \bibinfo {author} {\bibfnamefont {S.}~\bibnamefont {Basak}},\ }\bibfield  {title} {\bibinfo {title} {{Formation and decay of oscillons after inflation in the presence of an external coupling. Part I. Lattice simulations}},\ }\href {https://doi.org/10.1088/1475-7516/2024/10/082} {\bibfield  {journal} {\bibinfo  {journal} {JCAP}\ }\textbf {\bibinfo {volume} {10}},\ \bibinfo {pages} {082}},\ \Eprint {https://arxiv.org/abs/2406.00108} {arXiv:2406.00108 [hep-ph]} \BibitemShut {NoStop}%
\bibitem [{\citenamefont {Giblin}\ and\ \citenamefont {Tishue}(2019)}]{giblin2019preheating}%
  \BibitemOpen
  \bibfield  {author} {\bibinfo {author} {\bibfnamefont {J.~T.}\ \bibnamefont {Giblin}}\ and\ \bibinfo {author} {\bibfnamefont {A.~J.}\ \bibnamefont {Tishue}},\ }\bibfield  {title} {\bibinfo {title} {{Preheating in Full General Relativity}},\ }\href {https://doi.org/10.1103/PhysRevD.100.063543} {\bibfield  {journal} {\bibinfo  {journal} {Phys. Rev. D}\ }\textbf {\bibinfo {volume} {100}},\ \bibinfo {pages} {063543} (\bibinfo {year} {2019})}\BibitemShut {NoStop}%
\bibitem [{\citenamefont {Figueroa}\ \emph {et~al.}(2021)\citenamefont {Figueroa}, \citenamefont {Florio}, \citenamefont {Torrenti},\ and\ \citenamefont {Valkenburg}}]{Figueroa:2020rrl}%
  \BibitemOpen
  \bibfield  {author} {\bibinfo {author} {\bibfnamefont {D.~G.}\ \bibnamefont {Figueroa}}, \bibinfo {author} {\bibfnamefont {A.}~\bibnamefont {Florio}}, \bibinfo {author} {\bibfnamefont {F.}~\bibnamefont {Torrenti}},\ and\ \bibinfo {author} {\bibfnamefont {W.}~\bibnamefont {Valkenburg}},\ }\bibfield  {title} {\bibinfo {title} {The art of simulating the early universe. part i. integration techniques and canonical cases},\ }\href@noop {} {\bibfield  {journal} {\bibinfo  {journal} {Journal of Cosmology and Astroparticle Physics}\ }\textbf {\bibinfo {volume} {2021}}\bibinfo  {number} { (04)},\ \bibinfo {pages} {035}}\BibitemShut {NoStop}%
\bibitem [{\citenamefont {Pr{\"u}fer}\ \emph {et~al.}(2018)\citenamefont {Pr{\"u}fer}, \citenamefont {Kunkel}, \citenamefont {Strobel}, \citenamefont {Lannig}, \citenamefont {Linnemann}, \citenamefont {Schmied}, \citenamefont {Berges}, \citenamefont {Gasenzer},\ and\ \citenamefont {Oberthaler}}]{prufer2018observation}%
  \BibitemOpen
\bibfield  {number} {  }\bibfield  {author} {\bibinfo {author} {\bibfnamefont {M.}~\bibnamefont {Pr{\"u}fer}}, \bibinfo {author} {\bibfnamefont {P.}~\bibnamefont {Kunkel}}, \bibinfo {author} {\bibfnamefont {H.}~\bibnamefont {Strobel}}, \bibinfo {author} {\bibfnamefont {S.}~\bibnamefont {Lannig}}, \bibinfo {author} {\bibfnamefont {D.}~\bibnamefont {Linnemann}}, \bibinfo {author} {\bibfnamefont {C.-M.}\ \bibnamefont {Schmied}}, \bibinfo {author} {\bibfnamefont {J.}~\bibnamefont {Berges}}, \bibinfo {author} {\bibfnamefont {T.}~\bibnamefont {Gasenzer}},\ and\ \bibinfo {author} {\bibfnamefont {M.~K.}\ \bibnamefont {Oberthaler}},\ }\bibfield  {title} {\bibinfo {title} {Observation of universal dynamics in a spinor bose gas far from equilibrium},\ }\href@noop {} {\bibfield  {journal} {\bibinfo  {journal} {Nature}\ }\textbf {\bibinfo {volume} {563}},\ \bibinfo {pages} {217} (\bibinfo {year} {2018})}\BibitemShut {NoStop}%
\bibitem [{\citenamefont {Erne}\ \emph {et~al.}(2018)\citenamefont {Erne}, \citenamefont {Bücker}, \citenamefont {Gasenzer}, \citenamefont {Berges},\ and\ \citenamefont {Schmiedmayer}}]{Erne_2018}%
  \BibitemOpen
  \bibfield  {author} {\bibinfo {author} {\bibfnamefont {S.}~\bibnamefont {Erne}}, \bibinfo {author} {\bibfnamefont {R.}~\bibnamefont {Bücker}}, \bibinfo {author} {\bibfnamefont {T.}~\bibnamefont {Gasenzer}}, \bibinfo {author} {\bibfnamefont {J.}~\bibnamefont {Berges}},\ and\ \bibinfo {author} {\bibfnamefont {J.}~\bibnamefont {Schmiedmayer}},\ }\bibfield  {title} {\bibinfo {title} {Universal dynamics in an isolated one-dimensional bose gas far from equilibrium},\ }\href {https://doi.org/10.1038/s41586-018-0667-0} {\bibfield  {journal} {\bibinfo  {journal} {Nature}\ }\textbf {\bibinfo {volume} {563}},\ \bibinfo {pages} {225} (\bibinfo {year} {2018})}\BibitemShut {NoStop}%
\bibitem [{\citenamefont {Gazo}\ \emph {et~al.}(2023)\citenamefont {Gazo}, \citenamefont {Karailiev}, \citenamefont {Satoor}, \citenamefont {Eigen}, \citenamefont {Ga{\l}ka},\ and\ \citenamefont {Hadzibabic}}]{gazo2023universal}%
  \BibitemOpen
  \bibfield  {author} {\bibinfo {author} {\bibfnamefont {M.}~\bibnamefont {Gazo}}, \bibinfo {author} {\bibfnamefont {A.}~\bibnamefont {Karailiev}}, \bibinfo {author} {\bibfnamefont {T.}~\bibnamefont {Satoor}}, \bibinfo {author} {\bibfnamefont {C.}~\bibnamefont {Eigen}}, \bibinfo {author} {\bibfnamefont {M.}~\bibnamefont {Ga{\l}ka}},\ and\ \bibinfo {author} {\bibfnamefont {Z.}~\bibnamefont {Hadzibabic}},\ }\bibfield  {title} {\bibinfo {title} {Universal coarsening in a homogeneous two-dimensional bose gas},\ }\href@noop {} {\bibfield  {journal} {\bibinfo  {journal} {arXiv preprint arXiv:2312.09248}\ } (\bibinfo {year} {2023})}\BibitemShut {NoStop}%
\bibitem [{\citenamefont {Martirosyan}\ \emph {et~al.}(2024)\citenamefont {Martirosyan}, \citenamefont {Gazo}, \citenamefont {Etrych}, \citenamefont {Fischer}, \citenamefont {Morris}, \citenamefont {Ho}, \citenamefont {Eigen},\ and\ \citenamefont {Hadzibabic}}]{martirosyan2024universal}%
  \BibitemOpen
  \bibfield  {author} {\bibinfo {author} {\bibfnamefont {G.}~\bibnamefont {Martirosyan}}, \bibinfo {author} {\bibfnamefont {M.}~\bibnamefont {Gazo}}, \bibinfo {author} {\bibfnamefont {J.}~\bibnamefont {Etrych}}, \bibinfo {author} {\bibfnamefont {S.~M.}\ \bibnamefont {Fischer}}, \bibinfo {author} {\bibfnamefont {S.~J.}\ \bibnamefont {Morris}}, \bibinfo {author} {\bibfnamefont {C.~J.}\ \bibnamefont {Ho}}, \bibinfo {author} {\bibfnamefont {C.}~\bibnamefont {Eigen}},\ and\ \bibinfo {author} {\bibfnamefont {Z.}~\bibnamefont {Hadzibabic}},\ }\bibfield  {title} {\bibinfo {title} {A universal speed limit for spreading of quantum coherence},\ }\href@noop {} {\bibfield  {journal} {\bibinfo  {journal} {arXiv preprint arXiv:2410.08204}\ } (\bibinfo {year} {2024})}\BibitemShut {NoStop}%
\bibitem [{\citenamefont {Miles}(1967)}]{miles1967surface}%
  \BibitemOpen
  \bibfield  {author} {\bibinfo {author} {\bibfnamefont {J.~W.}\ \bibnamefont {Miles}},\ }\bibfield  {title} {\bibinfo {title} {Surface-wave damping in closed basins},\ }\href@noop {} {\bibfield  {journal} {\bibinfo  {journal} {Proceedings of the Royal Society of London. Series A. Mathematical and Physical Sciences}\ }\textbf {\bibinfo {volume} {297}},\ \bibinfo {pages} {459} (\bibinfo {year} {1967})}\BibitemShut {NoStop}%
\bibitem [{\citenamefont {Kidambi}(2009)}]{kidambi2009capillary}%
  \BibitemOpen
  \bibfield  {author} {\bibinfo {author} {\bibfnamefont {R.}~\bibnamefont {Kidambi}},\ }\bibfield  {title} {\bibinfo {title} {Capillary damping of inviscid surface waves in a circular cylinder},\ }\href@noop {} {\bibfield  {journal} {\bibinfo  {journal} {Journal of fluid mechanics}\ }\textbf {\bibinfo {volume} {627}},\ \bibinfo {pages} {323} (\bibinfo {year} {2009})}\BibitemShut {NoStop}%
\bibitem [{\citenamefont {for the Advancement~of Science}\ \emph {et~al.}(1952)\citenamefont {for the Advancement~of Science} \emph {et~al.}}]{british1952bessel}%
  \BibitemOpen
  \bibfield  {author} {\bibinfo {author} {\bibfnamefont {B.~A.}\ \bibnamefont {for the Advancement~of Science}} \emph {et~al.},\ }\href@noop {} {\bibinfo {title} {Bessel functions. part ii. functions of positive integer order}} (\bibinfo {year} {1952})\BibitemShut {NoStop}%
\bibitem [{\citenamefont {Zhang}\ \emph {et~al.}(2023)\citenamefont {Zhang}, \citenamefont {Borthwick},\ and\ \citenamefont {Lin}}]{zhang2023pattern}%
  \BibitemOpen
  \bibfield  {author} {\bibinfo {author} {\bibfnamefont {S.}~\bibnamefont {Zhang}}, \bibinfo {author} {\bibfnamefont {A.~G.}\ \bibnamefont {Borthwick}},\ and\ \bibinfo {author} {\bibfnamefont {Z.}~\bibnamefont {Lin}},\ }\bibfield  {title} {\bibinfo {title} {Pattern evolution and modal decomposition of faraday waves in a brimful cylinder},\ }\href@noop {} {\bibfield  {journal} {\bibinfo  {journal} {Journal of Fluid Mechanics}\ }\textbf {\bibinfo {volume} {974}},\ \bibinfo {pages} {A56} (\bibinfo {year} {2023})}\BibitemShut {NoStop}%
\bibitem [{\citenamefont {Guizar-Sicairos}\ and\ \citenamefont {Guti{\'e}rrez-Vega}(2004)}]{guizar2004computation}%
  \BibitemOpen
  \bibfield  {author} {\bibinfo {author} {\bibfnamefont {M.}~\bibnamefont {Guizar-Sicairos}}\ and\ \bibinfo {author} {\bibfnamefont {J.~C.}\ \bibnamefont {Guti{\'e}rrez-Vega}},\ }\bibfield  {title} {\bibinfo {title} {Computation of quasi-discrete hankel transforms of integer order for propagating optical wave fields},\ }\href@noop {} {\bibfield  {journal} {\bibinfo  {journal} {Josa A}\ }\textbf {\bibinfo {volume} {21}},\ \bibinfo {pages} {53} (\bibinfo {year} {2004})}\BibitemShut {NoStop}%
\bibitem [{\citenamefont {Gardiner}(2004)}]{gardiner2004handbook}%
  \BibitemOpen
  \bibfield  {author} {\bibinfo {author} {\bibfnamefont {C.~W.}\ \bibnamefont {Gardiner}},\ }\href@noop {} {\emph {\bibinfo {title} {Handbook of stochastic methods for physics, chemistry and the natural sciences}}},\ \bibinfo {edition} {3rd}\ ed.,\ \bibinfo {series} {Springer Series in Synergetics}, Vol.~\bibinfo {volume} {13}\ (\bibinfo  {publisher} {Springer-Verlag},\ \bibinfo {address} {Berlin},\ \bibinfo {year} {2004})\ pp.\ \bibinfo {pages} {xviii+415}\BibitemShut {NoStop}%
\bibitem [{\citenamefont {Bezrukov}\ and\ \citenamefont {Shaposhnikov}(2008)}]{Bezrukov:2007ep}%
  \BibitemOpen
  \bibfield  {author} {\bibinfo {author} {\bibfnamefont {F.~L.}\ \bibnamefont {Bezrukov}}\ and\ \bibinfo {author} {\bibfnamefont {M.}~\bibnamefont {Shaposhnikov}},\ }\bibfield  {title} {\bibinfo {title} {{The Standard Model Higgs boson as the inflaton}},\ }\href {https://doi.org/10.1016/j.physletb.2007.11.072} {\bibfield  {journal} {\bibinfo  {journal} {Phys. Lett. B}\ }\textbf {\bibinfo {volume} {659}},\ \bibinfo {pages} {703} (\bibinfo {year} {2008})}\BibitemShut {NoStop}%
\bibitem [{\citenamefont {Rubio}(2019)}]{Rubio:2018ogq}%
  \BibitemOpen
  \bibfield  {author} {\bibinfo {author} {\bibfnamefont {J.}~\bibnamefont {Rubio}},\ }\bibfield  {title} {\bibinfo {title} {{Higgs inflation}},\ }\href {https://doi.org/10.3389/fspas.2018.00050} {\bibfield  {journal} {\bibinfo  {journal} {Front. Astron. Space Sci.}\ }\textbf {\bibinfo {volume} {5}},\ \bibinfo {pages} {50} (\bibinfo {year} {2019})}\BibitemShut {NoStop}%
\bibitem [{\citenamefont {Kallosh}\ \emph {et~al.}(2013)\citenamefont {Kallosh}, \citenamefont {Linde},\ and\ \citenamefont {Roest}}]{Kallosh:2013yoa}%
  \BibitemOpen
  \bibfield  {author} {\bibinfo {author} {\bibfnamefont {R.}~\bibnamefont {Kallosh}}, \bibinfo {author} {\bibfnamefont {A.}~\bibnamefont {Linde}},\ and\ \bibinfo {author} {\bibfnamefont {D.}~\bibnamefont {Roest}},\ }\bibfield  {title} {\bibinfo {title} {Superconformal inflationary $\alpha$-attractors},\ }\href@noop {} {\bibfield  {journal} {\bibinfo  {journal} {Journal of High Energy Physics}\ }\textbf {\bibinfo {volume} {2013}},\ \bibinfo {pages} {1} (\bibinfo {year} {2013})}\BibitemShut {NoStop}%
\bibitem [{\citenamefont {Kaiser}\ and\ \citenamefont {Sfakianakis}(2014)}]{Kaiser:2013sna}%
  \BibitemOpen
  \bibfield  {author} {\bibinfo {author} {\bibfnamefont {D.~I.}\ \bibnamefont {Kaiser}}\ and\ \bibinfo {author} {\bibfnamefont {E.~I.}\ \bibnamefont {Sfakianakis}},\ }\bibfield  {title} {\bibinfo {title} {{Multifield Inflation after Planck: The Case for Nonminimal Couplings}},\ }\href {https://doi.org/10.1103/PhysRevLett.112.011302} {\bibfield  {journal} {\bibinfo  {journal} {Phys. Rev. Lett.}\ }\textbf {\bibinfo {volume} {112}},\ \bibinfo {pages} {011302} (\bibinfo {year} {2014})}\BibitemShut {NoStop}%
\bibitem [{\citenamefont {DeCross}\ \emph {et~al.}(2018{\natexlab{a}})\citenamefont {DeCross}, \citenamefont {Kaiser}, \citenamefont {Prabhu}, \citenamefont {Prescod-Weinstein},\ and\ \citenamefont {Sfakianakis}}]{DeCross:2015uza}%
  \BibitemOpen
  \bibfield  {author} {\bibinfo {author} {\bibfnamefont {M.~P.}\ \bibnamefont {DeCross}}, \bibinfo {author} {\bibfnamefont {D.~I.}\ \bibnamefont {Kaiser}}, \bibinfo {author} {\bibfnamefont {A.}~\bibnamefont {Prabhu}}, \bibinfo {author} {\bibfnamefont {C.}~\bibnamefont {Prescod-Weinstein}},\ and\ \bibinfo {author} {\bibfnamefont {E.~I.}\ \bibnamefont {Sfakianakis}},\ }\bibfield  {title} {\bibinfo {title} {{Preheating after Multifield Inflation with Nonminimal Couplings, I: Covariant Formalism and Attractor Behavior}},\ }\href {https://doi.org/10.1103/PhysRevD.97.023526} {\bibfield  {journal} {\bibinfo  {journal} {Phys. Rev. D}\ }\textbf {\bibinfo {volume} {97}},\ \bibinfo {pages} {023526} (\bibinfo {year} {2018}{\natexlab{a}})}\BibitemShut {NoStop}%
\bibitem [{\citenamefont {DeCross}\ \emph {et~al.}(2018{\natexlab{b}})\citenamefont {DeCross}, \citenamefont {Kaiser}, \citenamefont {Prabhu}, \citenamefont {Prescod-Weinstein},\ and\ \citenamefont {Sfakianakis}}]{DeCross:2016fdz}%
  \BibitemOpen
  \bibfield  {author} {\bibinfo {author} {\bibfnamefont {M.~P.}\ \bibnamefont {DeCross}}, \bibinfo {author} {\bibfnamefont {D.~I.}\ \bibnamefont {Kaiser}}, \bibinfo {author} {\bibfnamefont {A.}~\bibnamefont {Prabhu}}, \bibinfo {author} {\bibfnamefont {C.}~\bibnamefont {Prescod-Weinstein}},\ and\ \bibinfo {author} {\bibfnamefont {E.~I.}\ \bibnamefont {Sfakianakis}},\ }\bibfield  {title} {\bibinfo {title} {{Preheating after multifield inflation with nonminimal couplings, II: Resonance Structure}},\ }\href {https://doi.org/10.1103/PhysRevD.97.023527} {\bibfield  {journal} {\bibinfo  {journal} {Phys. Rev. D}\ }\textbf {\bibinfo {volume} {97}},\ \bibinfo {pages} {023527} (\bibinfo {year} {2018}{\natexlab{b}})}\BibitemShut {NoStop}%
\bibitem [{\citenamefont {DeCross}\ \emph {et~al.}(2018{\natexlab{c}})\citenamefont {DeCross}, \citenamefont {Kaiser}, \citenamefont {Prabhu}, \citenamefont {Prescod-Weinstein},\ and\ \citenamefont {Sfakianakis}}]{DeCross:2016cbs}%
  \BibitemOpen
  \bibfield  {author} {\bibinfo {author} {\bibfnamefont {M.~P.}\ \bibnamefont {DeCross}}, \bibinfo {author} {\bibfnamefont {D.~I.}\ \bibnamefont {Kaiser}}, \bibinfo {author} {\bibfnamefont {A.}~\bibnamefont {Prabhu}}, \bibinfo {author} {\bibfnamefont {C.}~\bibnamefont {Prescod-Weinstein}},\ and\ \bibinfo {author} {\bibfnamefont {E.~I.}\ \bibnamefont {Sfakianakis}},\ }\bibfield  {title} {\bibinfo {title} {{Preheating after multifield inflation with nonminimal couplings, III: Dynamical spacetime results}},\ }\href {https://doi.org/10.1103/PhysRevD.97.023528} {\bibfield  {journal} {\bibinfo  {journal} {Phys. Rev. D}\ }\textbf {\bibinfo {volume} {97}},\ \bibinfo {pages} {023528} (\bibinfo {year} {2018}{\natexlab{c}})}\BibitemShut {NoStop}%
\bibitem [{\citenamefont {Sfakianakis}\ and\ \citenamefont {van~de Vis}(2019)}]{Sfakianakis:2018lzf}%
  \BibitemOpen
  \bibfield  {author} {\bibinfo {author} {\bibfnamefont {E.~I.}\ \bibnamefont {Sfakianakis}}\ and\ \bibinfo {author} {\bibfnamefont {J.}~\bibnamefont {van~de Vis}},\ }\bibfield  {title} {\bibinfo {title} {{Preheating after Higgs Inflation: Self-Resonance and Gauge boson production}},\ }\href {https://doi.org/10.1103/PhysRevD.99.083519} {\bibfield  {journal} {\bibinfo  {journal} {Phys. Rev. D}\ }\textbf {\bibinfo {volume} {99}},\ \bibinfo {pages} {083519} (\bibinfo {year} {2019})}\BibitemShut {NoStop}%
\bibitem [{\citenamefont {Iarygina}\ \emph {et~al.}(2019)\citenamefont {Iarygina}, \citenamefont {Sfakianakis}, \citenamefont {Wang},\ and\ \citenamefont {Ach{\'u}carro}}]{Iarygina:2018kee}%
  \BibitemOpen
  \bibfield  {author} {\bibinfo {author} {\bibfnamefont {O.}~\bibnamefont {Iarygina}}, \bibinfo {author} {\bibfnamefont {E.~I.}\ \bibnamefont {Sfakianakis}}, \bibinfo {author} {\bibfnamefont {D.-G.}\ \bibnamefont {Wang}},\ and\ \bibinfo {author} {\bibfnamefont {A.}~\bibnamefont {Ach{\'u}carro}},\ }\bibfield  {title} {\bibinfo {title} {Universality and scaling in multi-field $\alpha$-attractor preheating},\ }\href@noop {} {\bibfield  {journal} {\bibinfo  {journal} {Journal of Cosmology and Astroparticle Physics}\ }\textbf {\bibinfo {volume} {2019}}\bibinfo  {number} { (06)},\ \bibinfo {pages} {027}}\BibitemShut {NoStop}%
\bibitem [{\citenamefont {Rayleigh}(1873)}]{rayleigh1873some}%
  \BibitemOpen
\bibfield  {number} {  }\bibfield  {author} {\bibinfo {author} {\bibfnamefont {J.~W. S.~B.}\ \bibnamefont {Rayleigh}},\ }\href@noop {} {\emph {\bibinfo {title} {Some general theorems relating to vibrations}}}\ (\bibinfo  {publisher} {London Mathematical Society},\ \bibinfo {year} {1873})\BibitemShut {NoStop}%
\bibitem [{\citenamefont {Herreman}\ \emph {et~al.}(2019)\citenamefont {Herreman}, \citenamefont {Nore}, \citenamefont {Guermond}, \citenamefont {Cappanera}, \citenamefont {Weber},\ and\ \citenamefont {Horstmann}}]{Herreman2019}%
  \BibitemOpen
  \bibfield  {author} {\bibinfo {author} {\bibfnamefont {W.}~\bibnamefont {Herreman}}, \bibinfo {author} {\bibfnamefont {C.}~\bibnamefont {Nore}}, \bibinfo {author} {\bibfnamefont {J.-L.}\ \bibnamefont {Guermond}}, \bibinfo {author} {\bibfnamefont {L.}~\bibnamefont {Cappanera}}, \bibinfo {author} {\bibfnamefont {N.}~\bibnamefont {Weber}},\ and\ \bibinfo {author} {\bibfnamefont {G.~M.}\ \bibnamefont {Horstmann}},\ }\bibfield  {title} {\bibinfo {title} {Perturbation theory for metal pad roll instability in cylindrical reduction cells},\ }\href {https://doi.org/10.1017/jfm.2019.642} {\bibfield  {journal} {\bibinfo  {journal} {Journal of Fluid Mechanics}\ }\textbf {\bibinfo {volume} {878}},\ \bibinfo {pages} {598–646} (\bibinfo {year} {2019})}\BibitemShut {NoStop}%
\bibitem [{\citenamefont {Bassett}\ \emph {et~al.}(1999)\citenamefont {Bassett}, \citenamefont {Tamburini}, \citenamefont {Kaiser},\ and\ \citenamefont {Maartens}}]{bassett1999metric}%
  \BibitemOpen
  \bibfield  {author} {\bibinfo {author} {\bibfnamefont {B.~A.}\ \bibnamefont {Bassett}}, \bibinfo {author} {\bibfnamefont {F.}~\bibnamefont {Tamburini}}, \bibinfo {author} {\bibfnamefont {D.~I.}\ \bibnamefont {Kaiser}},\ and\ \bibinfo {author} {\bibfnamefont {R.}~\bibnamefont {Maartens}},\ }\bibfield  {title} {\bibinfo {title} {Metric preheating and limitations of linearized gravity},\ }\href@noop {} {\bibfield  {journal} {\bibinfo  {journal} {Nuclear Physics B}\ }\textbf {\bibinfo {volume} {561}},\ \bibinfo {pages} {188} (\bibinfo {year} {1999})}\BibitemShut {NoStop}%
\bibitem [{\citenamefont {Berges}\ \emph {et~al.}(2004)\citenamefont {Berges}, \citenamefont {Bors\'anyi},\ and\ \citenamefont {Wetterich}}]{PhysRevLett.93.142002}%
  \BibitemOpen
  \bibfield  {author} {\bibinfo {author} {\bibfnamefont {J.}~\bibnamefont {Berges}}, \bibinfo {author} {\bibfnamefont {S.}~\bibnamefont {Bors\'anyi}},\ and\ \bibinfo {author} {\bibfnamefont {C.}~\bibnamefont {Wetterich}},\ }\bibfield  {title} {\bibinfo {title} {Prethermalization},\ }\href {https://doi.org/10.1103/PhysRevLett.93.142002} {\bibfield  {journal} {\bibinfo  {journal} {Phys. Rev. Lett.}\ }\textbf {\bibinfo {volume} {93}},\ \bibinfo {pages} {142002} (\bibinfo {year} {2004})}\BibitemShut {NoStop}%
\end{thebibliography}%


\section*{Methods}
\setcounter{section}{0}
\setcounter{figure}{0}
\renewcommand{\tablename}{\textbf{Extended Data Table}}
\renewcommand{\figurename}{\textbf{Extended Data Figure}}
\renewcommand{\thefigure}{\textbf{\arabic{figure}}}

\subsection{Experimental setup}\label{app::ExpSetup}

Our experimental system is a stratified biphasic liquid solution sealed within a cylindrical vessel, with radius $\ro=\SI{40}{\milli\meter}$ and height $2h_0=\SI{30}{\milli\meter}$. The hard radial walls are made from machined nylon, and the ceiling and floor are made of transparent glass. The lower fluid is an aqueous solution of potassium carbonate and distilled water with density $\rho_{\mathrm{b}}=\SI{1288.7\pm2.3}{\kilogram\per\meter^3}$ and kinematic viscosity $\nu_{\mathrm{b}}=\SI{2.35(2)}{\milli\meter^2\per\second}$, while the top fluid is an organic solution of ethanol and distilled water with density $\rho_{\mathrm{t}}=\SI{920.6\pm2.3}{\kilogram\per\meter^3}$ and viscosity $\nu_{\mathrm{t}}=\SI{3.40(2)}{\milli\meter^2\per\second}$. These fluids are immiscible and separate to form an interface with interfacial tension $\sigma=\SI{3.0(5)}{\milli\newton\per\meter}$. Since the vessel is filled with equal volumes of each phase, the interface at rest lies a distance $h_0=\SI{15}{\milli\meter}$ from both ceiling and floor.  

The vessel is mounted on a bespoke shaking platform, suspended by springs from a base securely mounted on an optical table to reduce vibrational noise. The platform is sinusoidally driven by a voice coil actuator and vertically glides along pneumatic air bearings to reduce off-axis oscillations. The typical ratio of horizontal to vertical acceleration is $|a_\perp|/|a_z|=\num{4.3e-3}$ for the datasets presented here. 

A checkerboard pattern is placed underneath the fluid vessel and recorded by a Phantom$^\text{®}$ VEO 640 camera at 100 frames per second. Waves on the interface distort the recorded pattern and their three-dimensional profile $\xi(t,r,\theta)$ is reconstructed via Fourier transform profilometry~\cite{Wildeman2018Real-timeBackdrop}. By reconstructing a series of undistorted patterns, we estimate that the resolution of our implementation is in the order of $\sim \SI{1}{\micro\meter}$.

Data acquisition is performed with a National Instruments DAQ PCIeX card, which operates the actuator used to drive the platform while synchronously triggering the camera's recording. For the present work, 25 experimental repetitions were performed, each with 350 cycles of the driver, and the interfacial profile $\xi(t,r,\theta)$ was reconstructed throughout. Across this ensemble of runs, the unstable growth of interface waves stagnates at different times. We remedy this varying time delay on the interfacial evolution by synchronising all runs and ensuring the largest wave reaches the fully nonlinear stage at $t_{\text{sync}} \sim\SI{32}{\second}$ across all repetitions.

\subsection{Decomposition of interfacial waves}\label{app::WettingCondition}
The interface height $\xi(t,r,\theta)$ may be written in terms of the spatial profiles $\Psi_{mn}(r,\theta)$, which are eigenfunctions of the $2$D Laplacian in a disk of radius $r=r_0$ with corresponding eigenvalues $k_{mn}$. By writing the interface as 
\begin{equation}
    \xi(t,r,\theta)=\sum_{m=-\infty}^{\infty}\sum_{n=0}^{\infty}\xi_{m,n}(t)\Psi_{m,n}(r,\theta),
\end{equation}
we must require $\xi_{-m,n}=\xi_{m,n}^*$ to ensure the total field $\xi(t,r,\theta)$ remains a real-valued quantity, while the amplitudes $\xi_{mn}(t)$ are complex-valued. The spatial eigenfunctions are separable in radius and angle, and may be written as $\Psi_{mn}(r,\theta)=R_{mn}(r)\exp(im\theta)$, where the radial component $R_{mn}(r)$ satisfies Bessel's equation of order $m$ and its form is determined by the choice of boundary condition at $r=r_0$.

We approximate the radial boundary conditions on the fluid-fluid interface by free, or Neumann, boundary conditions. This corresponds to an assumed freely slipping contact line of the interface with the outer wall. However, the capillary properties between the fluids and the wall material determine the precise contact line's motion and cause a meniscus to form~\cite{miles1967surface}. Due to the low interfacial tension and the smooth wall, these effects are reduced in our experimental implementation. However, the full radial boundary condition is more accurately captured by a ``wetting condition'' in which the contact line's velocity is proportional to the contact angle with the wall~\cite{kidambi2009capillary}. As a result, each radial eigenmode $R_{mn}(r)$ experiences a different Robin boundary condition, $R_{m n}(r) + \lambda_{m n} R_{m n}'(r)$, where the Robin constant $\lambda_{mn}$ is inversely proportional to the (complex) frequency $\Omega_{m n}$ of that mode, i.e., $\lambda_{m n} \propto \Omega_{m n}^{-1}$.
 
While low-frequency modes obey a condition close to free (Neumann), this assumption breaks down in the limit of high frequency where the solutions tend towards that of fixed (Dirichlet) boundary conditions. However, since higher frequencies imply larger wavenumber $k$, the boundary condition matters less for high $k$ as the solutions for Neumann and Dirichlet asymptotically approach each other~\cite{british1952bessel}, and so our approximation of Neumann boundary conditions is well justified.
 The resulting radial eigenfunctions are $R_{mn}\equiv\mathcal{N}_{mn}J_{|m|}\left(k_{mn}r\right)$, with $J_{|m|}'(k_{mn}r_0)=0$ and normalisation constant
\begin{equation}
    \mathcal{N}_{mn}^{-2} =\left(1-\frac{m^2}{k_{mn}^2r_0^2}\right) J_{|m|}^2(k_{mn} r_0),
\end{equation}
ensuring the orthonormality of the radial basis. Here, $J_{|m|}$ is the $m$-th order Bessel function of the first kind, and the non-negative integer $n$ labels the order of the zeros of $J_{|m|}'$.

\subsection{Analysis of observed interface }\label{appendix:ModeExtraction}
Instantaneous mode amplitudes $\xi_{mn}(t)$ are obtained by decomposing the reconstructed interfacial height $\xi$ on the basis of theoretical spatial eigenfunctions~$\Psi_{m,n}(r,\theta)$, in a similar fashion to recently developed modal decomposition methodology~\cite{zhang2023pattern}. First, a fast Fourier transform (FFT) is performed over the angle of the height profile so that the data is now expressed in terms of the azimuthal number $m$, that is, $\xi(t,r,\theta)\to\xi_m(t,r)$. Then, this quantity is projected onto the basis of radial functions $R_{mn}$ via a finite-domain Hankel transform~\cite{guizar2004computation} for zeros of $J_{m}'$ to obtain the quantity $\xi_{mn}(t)$. We then compute the root mean squared (rms) value of this oscillating amplitude over time intervals of $T_0\equiv 1/f_0$ (period of the primary mode) to obtain the slow-time envelopes $\bar{\xi}_{mn}(t)$,
\begin{equation}
\label{eq:RMS_amplitude_def}
    \bar{\xi}_{mn}(t) = \sqrt{\frac{1}{T_0}\int_{t}^{t+T_0}\mathrm{d}t~|\xi_{mn}(t)|^2}\,.
\end{equation}
We employ a short-time FFT on $\xi$ and $\xi_m$ to retrieve time-dependent frequency amplitudes $\xi_{\omega}(t,r,\theta)$ and $\xi_{m,\omega}(t,r)$, used in the computation of correlation functions.

\begin{figure*}[ht]
\centering
\includegraphics[width=\linewidth]{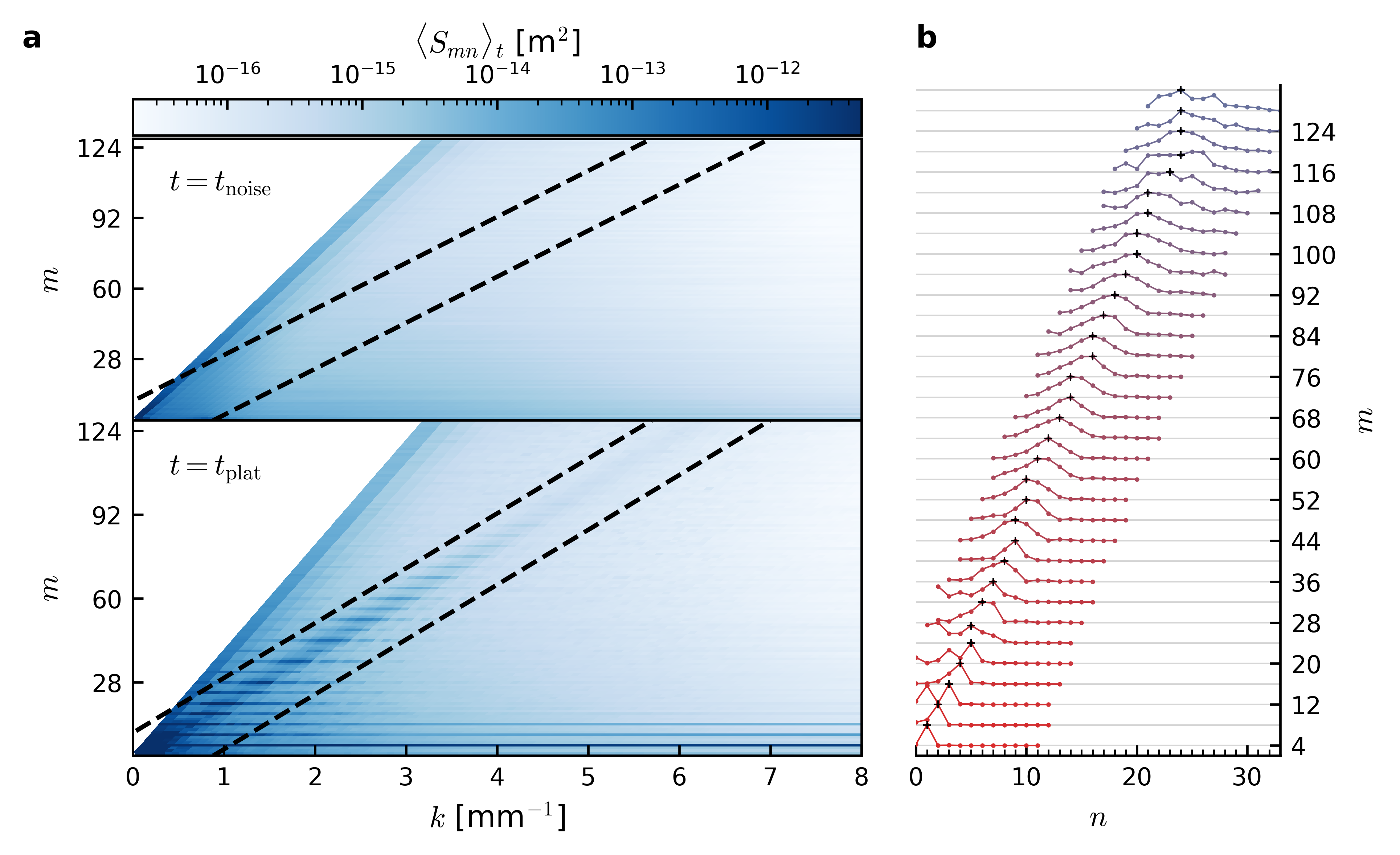}
\caption{\textbf{Hankel transformed spectral data used to locate the $(m,n)$ indices of excited modes.} 
Panel \textbf{a} displays two spectra, both averaged over the same ensemble of 25 experimental runs, but also averaged over two seperate intervals of time $t_{\mathrm{noise}} = [\SI{0}{\second},\SI{5}{\second}]$ and $t_{\mathrm{plat}} = [\SI{38.76}{\second}, \SI{50.35}{\second}]$. The upper heatmap is the spectrum of noise present before any modes are excited, the lower is the spectrum with all modes maximally excited.
In panel \textbf{b}, the profile of the excited spectrum with background noise removed is plotted for each azimuthal number that becomes excited (multiples of 4). Each horizontal band displays $\langle S_{mn}\rangle_{t_{\mathrm{plat}}}-\langle S_{mn}\rangle_{t_{\mathrm{noise}}}$ in arbitrary units at a fixed $m$ against values of $n$ corresponding to those within the dashed black lines of panel a. The cross-hairs mark peaks of greatest prominence corresponding to an excited $(m,n)$ mode.
}
\label{fig:HankelPeaks}
\end{figure*}

Power spectral densities (PSD), $\langle S_{mn}\rangle_{t}$, at two different times are computed by averaging $|\xi_{mn}(t)|^2$ across the synchronised ensemble of 25 experimental repetitions, and then averaging over two separate time intervals: $t_{\mathrm{noise}} = [\SI{0}{\second},\SI{5}{\second}]$ and $t_{\mathrm{plat}} = [\SI{38.76}{\second}, \SI{50.35}{\second}]$. During $t_{\mathrm{noise}}$, no wave modes have been excited or resolved above the amplitude threshold for detection, and so the spectrum averaged over this period is background noise primarily due to the limitations of the detection scheme. The period $t_{\mathrm{plat}}$ corresponds to 36 full cycles of $\omega_0$ during the steady-state where the amplitude of all excited modes are maximal. The two spectra are displayed in Extended Data Figure~\ref{fig:HankelPeaks}\textbf{a}. 

While $\langle S_{mn}\rangle_{t_{\mathrm{noise}}}$ displays a smooth gradient, peaks emerge in $\langle S_{mn}\rangle_{t_{\mathrm{plat}}}$ at azimuthal numbers that are multiples of the $m=4$ primary. These peaks correspond to excited modes. The background noise is removed from $\langle S_{mn}\rangle_{t_{\mathrm{plat}}}$ to better distinguish the location of these excited peaks, and for each excited $m$ we determine which $n$ corresponds to the most prominent peak. The profile of the excited mode peaks are presented in Extended Data Figure~\ref{fig:HankelPeaks}\textbf{b}. Each $(m,n)$ pair obtained by this approach can be fixed to retrieve the time-dependent amplitude $\xi_{m n}(t)$ (Fig~\ref{fig:one}\textbf{b}) of an eigenmode excited over the course of the experiment. This set of modes is then used to calculate the PSD, $S_{mn}=|\xi_{mn}|^2$, with corresponding wavenumber $k_{mn}$, displayed in Fig \ref{fig:one}\textbf{c-e}.

\subsection{Lagrangian Description}\label{appendix:Lagrangian}

We write the perturbative Lagrangian for interfacial wave-modes in terms of the effective scalar field $\xi_a$ as:
\begin{align}
\label{eqn:LagrTot}
    L=& \sum_{N\geq2}L^{(N)} =L^{(2)}+L^{(3)}+L^{(4)}+\cdots\notag
    \\=&\frac{1}{2}\sum_{a}\left[ c_a\left(|\dot\xi_a|^2-\omega^2_a|\xi_a|^2\right)+ a_z(t)\mathrm{At}|\xi_a|^2\right]\notag
    \\&+\frac{1}{2}\sum_{a,b,c}f_{cab} \xi_c\dot\xi_a\dot\xi_b\notag
    \\&+\frac{1}{4}\sum_{a,b,c,d}\left[g_{cdab} \dot\xi_a\dot\xi_b+h_{abcd}\xi_a\xi_b\right]\xi_c\xi_d\notag
    \\&+\cdots 
\end{align}
where 
\begin{subequations}
\label{eqn:coupling_const_definitions}
\begin{align}
c_a &=\frac{1}{k_a\tanh(k_a h_0)}, \\
f_{cab} &=\mathrm{At}\,\mathcal{A}_{cab}, \label{fdef} \\
g_{cdab} &=\mathcal{A}_{cdab},\\
h_{abcd} &=\frac{1}{2}\tilde{\sigma}\mathcal{B}_{abcd},
\label{cfghdef}
\end{align}
\end{subequations}
with $\mathrm{At}=\frac{\rho_{\mathrm{b}}-\rho_{\mathrm{t}}}{\rho_{\mathrm{b}}+\rho_{\mathrm{t}}}$, the Atwood number, and $\tilde{\sigma}=\frac{\sigma}{(\rho_{\mathrm{b}}+\rho_{\mathrm{t}})}$. An explicit expression of this weakly nonlinear Lagrangian is provided in \cite{Barroso_2023}. In the absence of a forcing term $a_z(t)$, the quadratic Lagrangian $L^{(2)}$ describes the free dynamics of modes oscillating at their dispersion frequency $\omega_a$, given by
\begin{equation}
    \omega_a^2 = \frac{g\mathrm{At}+\tilde{\sigma}k_a^2}{c_a}.
\end{equation}

The $N$-th order interactions between instabilities are represented by the terms $L^{(3)}$ and $L^{(4)}$, and their strength is indicated by the coefficients $\mathcal{A}_{cab}$, $\mathcal{A}_{dcab}$ and $\mathcal{B}_{abcd}$. 
Integrals across the spatial profiles of the modes are contained in the three nonlinear coefficients in Eq.~\eqref{eqn:LagrTot}, as first described in~\cite{miles1986NonlinearStratified}. For instance, the integral entering the cubic nonlinear coefficient reads
\begin{equation} \label{eqn:cubic_coeff_form}
    \mathbb{C}_{cab} = \frac{1}{\pi r_0^2}\int_0^{2\pi}\int_0^{r_0}\mathrm{d}\theta\,\mathrm{d}r\,r R_{c}R_{a}R_{b}e^{i(m_a+m_c+m_b)\theta},
\end{equation}
where $a=(m_a,n_a)$ and $R_a\equiv R_{m_an_a}$.
The angular component reveals a momentum conservation law across the azimuthal numbers of the modes involved in the interfacial interactions. Consequently, the complete cubic nonlinear coefficient may be expressed as
\begin{equation}
     \mathcal{A}_{cab}\propto \delta_{m_a+m_b+m_c,0}\underbrace{\int_0^1 dx\cdot x \cdot J_c(p_cx)J_a(p_ax)J_b(p_bx)}_\text{$\mathbf{A}_{cab}$},
\end{equation}
where $p_a\equiv p_{m_a n_a}$ are the dimensionless zeros of the derivatives of Bessel functions. The same logic applies to $\mathcal{A}_{cdab}$ and $\mathcal{B}_{abcd}$, where quartic integral terms $\mathbb{C}_{cab}$ and $\mathbb{D}_{cdab}$ are taken into consideration; for their complete form, see~\cite{Barroso_2023,miles1986NonlinearStratified}.

In this instance, we examine a dominant mode $a=(m,n)$ with amplitude substantially higher than any other wave $b=(m',n')$, $\xi_{mn}\gg \xi_{m'n'}$. We restrict our analysis to the nonlinear coefficients up to the third and fourth orders of self-mixing of $m$ in a first approximation:
\begin{align}
    A^{(3)}_{2m,n'} &\sim \delta_{2m+m',0}\left(2\mathbf{A}_{aab}+\mathbf{A}_{baa}\right),\\
    A^{(4)}_{3m,n'} &\sim \delta_{3m+m',0} \left(2\mathbf{A}_{aaab}-\mathbf{A}_{abbb}\right),
\end{align}
where the coefficients $A^{(N)}_{a}$ represent the strength for the cubic or quartic self-interactions of the primary. As can be seen in Fig. \ref{fig:few}\textbf{c}, where the modes $(8,2)_{2\omega_0}$ and $(12,3)_{3\omega_0}$ are observed, our experimental data accurately captures this theoretical result.

\subsection{Numerical Coefficients}\label{appendix:Coefficients}

By isolating the equations of motion for $(m,n)_{\omega}=(8,n')_{2\omega_0}$ and $(m,n)_{\omega_0}=(12,n'')_{3\omega_0}$, respectively, in Eq.~\eqref{eqn:LagrTot}, the numerical coefficients for the prevalent three- and four-wave scattering have been derived. The nonlinear coefficients linked to the emergence of the mode $(8,n')_{2\omega_0}$ are evaluated taking into account only the dominant mode's self interactions. This leads to:
\begin{equation}
    |A_{8,n'}|\sim
    [2\mathcal{A}_{448}-\frac{1}{2}\mathcal{A}_{844}].
\end{equation}
For the creation of the $(12,n'')_{3\omega_0}$ we consider the contribution of wave-mixing $(4,1)_{\omega_0}+(8,n')_{2\omega_0}$ in a cubic interaction together with the quartic process $(4,1)_{\omega_0}+(4,1)_{\omega_0}+(4,1)_{\omega_0}$. This is equivalent to evaluating the amplitude for $a=(12,n'')_{3\omega_0}$:
\begin{align}
    |A_{a}|&\sim
    \mathrm{At}|A_{4,1}||A_{8,2}|(6\mathcal{A}_{48a}+3\mathcal{A}_{84a}-2\mathcal{A}_{a84})
    \nonumber\\&+|A_{4,1}|^3\big[(\mathcal{A}_{444a}-\mathcal{A}_{a444})+\tilde{\sigma}\mathcal{B}_{444a}\big].
\end{align}
These computations have produced $n'=2$ and $n''=3$ as shown in Extended Data Table~\ref{table:coeff}, which are consistent with the experimental findings. The coefficients $|A_{4,1}|$ and $|A_{8,2}|$ are experimentally obtained by the wave-amplitudes in the nonlinear regime.

\subsection{Correlation Functions}\label{app:CorrFunct}

In order to describe the features of wave interactions in comparison with field theories, we introduce two correlation measures following~\cite{schweigler2017experimental}. For the analysis of individual interactions, we define 
\begin{equation}
    g_N(\xi_{1}\dots \xi_{N}) = \langle \xi_{1}\dots \xi_{N}\rangle,
\end{equation}
in which $\xi_{i}$ denotes the mode amplitudes for both azimuthal number and frequency fixed, i.e., $\xi_{i} \equiv \xi_{m,\omega}(t,r)$. Our analysis is focused on the connected part $g^{\text{con}}_N(\xi_{1}\dots \xi_{N})$ of the correlation functions. Specifically, in Fig.~\ref{fig:few}\textbf{c}, we consider $\bar{g}_N (t)$ as the normalised cumulant for individual interactions, defined as
\begin{equation}
\label{eqn:g_N}
    \bar g_N(t) \equiv \frac{|\langle \xi_{1}\dots \xi_{N}\rangle_c|}{{\langle|\xi_{1}\dots \xi_{N}|\rangle}}.
\end{equation}
The averages $\langle \cdot \rangle$ are calculated over $r$, $\theta$, and a statistical ensemble of 25 runs. The connected correlation function, or statistical cumulant, of the arguments is indicated here by $\langle \cdot \rangle_c$, while their full correlation function is indicated by $\langle \cdot \rangle$. The statistical cumulant $\langle \cdot \rangle_c$ is computed using the iterative procedure outlined in~\cite{gardiner2004handbook}.

On the other hand, in the deeply nonlinear regime, we examine the state of the interactions using frequency correlation functions
\begin{equation}
\label{eqn:G_N}
    G_N (\omega_1,\dots,\omega_N)= \langle \xi_{\omega_1}\dots \xi_{\omega_N}\rangle,
\end{equation}
where now the degrees of freedom are simply the time-dependent Fourier amplitudes $\xi_{\omega}(t,r,\theta)$.
This is equivalent to accounting for the sum of $g_N$ on all available interactions between $(N-2)$ modes vibrating at some $\omega_j$ and those vibrating at $\omega_{\textrm{in}},\omega_{\textrm{out}}$,
\begin{equation}
   \label{eqn:G_N_2} G_N(\omega_1,\dots,\omega_N)=\sum_{m_1,\dots,m_N}g_N(\xi_{m_1,\omega_1}\dots\xi_{m_N,\omega_N}).
\end{equation}
For Fig.~\ref{fig:many}, we consider its normalised form:
\begin{equation}
    \bar{G}_N(\omega_1,\dots,\omega_N) = \frac{|\langle \xi_{\omega_1}\dots \xi_{\omega_N}\rangle_c|}{{\langle|\xi_{\omega_1}\dots \xi_{\omega_N}|\rangle}},
\end{equation}
where we average across all dimensions $(t,r,\theta)$ and the ensemble of repetitions.

\subsection{Relationship to Cosmological Models}\label{app:cosmo}
For generic models of early-universe inflation, including those that provide the closest fit between theoretical predictions and high-precision measurements of anisotropies in the cosmic microwave background radiation \cite{Martin:2013tda,kaiser2016nonminimal}, one may write the Lagrangian density governing the matter degrees of freedom in the general form
\begin{equation}
   \frac{ {\cal L}}{\sqrt{-g}} =  \frac{1}{2} g^{\mu\nu} {\cal G}_{IJ} (\phi^K) \partial_\mu \phi^I \partial_\nu \phi^J - V (\phi^K) .
   \label{Lmultifield}
\end{equation}
Here $V (\phi^K)$ is the interaction potential among ${\cal N}$ interacting scalar fields ($I, J = 1, 2, 3, ... , \,{\cal N}$), and ${\cal G}_{IJ} (\phi^K)$ is the metric on the field-space manifold, which modifies the canonical kinetic terms and induces additional interactions among the ${\cal N}$ fields. To avoid instabilities, the components ${\cal G}_{IJ} (\phi^K)$ depend on the fields $\phi^K$ but not on their derivatives $\partial_\mu \phi^K$. For effectively single-field models, the dynamics during and after inflation depend only on the functions ${\cal G}_{IJ} (\phi^K) \rightarrow {\cal G} (\phi)$ and $V (\phi^K) \rightarrow V (\phi)$.

Following a conformal transformation to the so-called Einstein frame, the form in Eq.~(\ref{Lmultifield}) describes successful models such as Higgs inflation \cite{Bezrukov:2007ep,Rubio:2018ogq}, $\alpha$-attractors \cite{Kallosh:2013yoa}, and their many variants \cite{Kaiser:2013sna,kaiser2016nonminimal}, in which the scalar fields have nonminimal gravitational couplings to the spacetime Ricci scalar. Such nonminimal couplings are a generic feature for any self-interacting scalar field in a curved spacetime \cite{kaiser2016nonminimal}.  

To compare with the effective Lagrangian that governs the interfacial wave-modes, Eq.~(\ref{eqn:LagrTot}), we may consider single-field cosmological models and expand the real-valued scalar field $\phi (x^\mu)$ in the basis of spatial eigenmodes $Y_a (x)$,
\begin{equation}
\phi (x, t) = \sum_a  \xi_a (t) \, e^{-i \omega_a t} \, Y_a (x) \,.
\end{equation}
It is then clear that the nonlinear damping term $g_{cdab} \dot\xi_a\dot\xi_b\xi_c\xi_d$ in the effective (fluid) Lagrangian in Eq.~(\ref{eqn:LagrTot}) arises from a term of the form $\phi^2\dot\phi^2$ (corresponding to ${\cal G} (\phi)\propto \phi^2$ in the single-field version of Eq.~(\ref{Lmultifield})) while the quartic interaction term $h_{abcd}\xi_a\xi_b\xi_c\xi_d$ in Eq.~(\ref{eqn:LagrTot}) corresponds to $V(\phi)\propto \phi^4$ in Eq.~(\ref{Lmultifield}).           
These terms play a crucial role in inflationary models for which post-inflation reheating has been studied both semi-analytically 
\cite{DeCross:2015uza,DeCross:2016fdz,DeCross:2016cbs,Sfakianakis:2018lzf,Iarygina:2018kee} and numerically (via lattice simulations) \cite{nguyen2019nonlinear,van2020time}. Note that the term $f_{cab} \xi_c\dot\xi_a\dot\xi_b$ in Eq.~(\ref{eqn:LagrTot}) is not generically present in models of reheating, but this can be tuned to be small in the fluid experiment as it scales with the difference between the densities of the two fluids, Eq.~\ref{fdef}. When applied to a spatially flat Friedmann-Robertson-Lema\^{i}tre-Walker spacetime geometry - appropriate for the end of cosmic inflation - the derivative term in Eq.~(\ref{Lmultifield}) takes the form $(\partial_\mu \phi)^2 = \dot{\phi}^2 - a(t)^{-2}(\nabla \phi)^2 $.  

Once inflation has ended, the inflaton condensate oscillates quasi-periodically around the global minimum of its potential. These oscillations can drive resonant particle production in any field(s) that are coupled as in Eq.~(\ref{Lmultifield}) to the spatially homogeneous condensate, which manifests as rapidly growing amplitudes of higher-momentum modes of the fields $\phi^I (x^\mu)$ \cite{kofman1997towards,amin2015nonperturbative}. Referring back to the effective (fluid) Lagrangian in Eq.~(\ref{eqn:LagrTot}), the term $a_z(t)\mathrm{At}|\xi_a|^2$, which includes the oscillatory driving force in $a_z(t)$, arises in Eq.~(\ref{Lmultifield}) from $V(\phi)\propto \phi^4$ which, upon expanding the field as $\phi(x^\mu)=\varphi(t) + \delta\phi(x^\mu)$ around the oscillating field condensate $\varphi(t)$ during reheating, induces a term $\propto \varphi(t)^2 \delta\phi^2$. The same term can also arise in two-field models from a quartic interaction term $V(\phi,\chi)\propto \phi^2\chi^2$, which couples an oscillating (inflaton) field condensate $\chi(t)$ to the fluctuations of a matter field $\phi (x^\mu)$ undergoing parametric resonance. For an effectively single-field inflationary model, specified by the functions ${\cal G} (\phi)$ and $V(\phi)$, we may relate the amplitude $\varphi_0$ and frequency $\omega$ of the inflaton condensate's oscillations to couplings within the model. Thus, by varying the amplitude and frequency with which the fluids in the experiments described here are driven, we may then vary experimental conditions to explore the parameter space of a given cosmological model.     

\subsection{One Mode Reduced Dynamics}
To predict the amplitude of the primary mode during the far-from equilibrium steady-state, we use a reduced dynamics model in which only the behaviour of the primary mode $\xi_a$ is considered. By dropping all other modes from the Lagrangian Eq.~\ref{eqn:LagrTot} and truncating at quartic order, we obtain an effecting one mode reduced Lagrangian:
\begin{multline}
\label{eqn:OneModeLagrTot}
    L^{(1)}=
    \frac{1}{2} c_a(|\dot\xi_a|^2-\omega^2_a|\xi_a|^2)+ \frac{1}{2}a_z(t)\mathrm{At}|\xi_a|^2 \\
    +\frac{1}{4} g_{aaaa} \dot\xi_a^2 \xi_a^2+\frac{1}{4}h_{aaaa}\xi_a^4
\end{multline}
where the self-coupling coefficients are the sum of non-vanishing terms e.g. $g_{aaaa}=g_{ddDD}+g_{dDdD}+g_{dDDd}$ ($d/D$ denoting $(+m,n)/(-m,n)$), and likewise for $h_{aaaa}$.
Notice that there is no self-coupling at cubic order since the coefficient $f_{aaa}$ vanishes, which can be seen from setting $a=b=c$ in Eq.~\ref{eqn:cubic_coeff_form}.
The primary mode oscillates at $\omega_0$ with an amplitude $\overline{\xi}_a(t)$ that changes over a slower timescale, and so we expect it to take the form:
\begin{equation} \label{eqn:xi_form_axiom}
    \xi_a (t) = \frac{1}{\sqrt{2}} \left(\overline{\xi}_a (t) e^{i \omega_0 t} + \overline{\xi}_a^* (t) e^{-i \omega_0 t}\right)
\end{equation}
Following the same procedure as in~\cite{miles1984nonlinear}, the form of Eq.~\ref{eqn:xi_form_axiom} is substituted into the reduced Lagrangian ~\ref{eqn:OneModeLagrTot}, which is then integrated over a time period of $2\pi/\omega_0$ to remove oscillating quantities and leave terms only dependent on the slowly varying amplitude $\overline{\xi}_a (t)$. An effective equation of motion for this quantity is determined from the time averaged Lagrangian by use of the Euler-Lagrange equation, accounting for linear damping $\gamma$ with Raleigh’s
dissipation function~\cite{rayleigh1873some} $Q_0 = c_a \gamma_a \dot \xi_a^2$.
We solve for fixed-point solutions of the equation of motion by setting the time derivatives of $\overline{\xi}_a$ to zero, the nontrivial solution corresponds to the saturation amplitude of the primary mode which is Eq.~\ref{eqn:SatAmp} of the main text. The formula tells us that the saturation amplitude depends on the balance between driving acceleration, damping rate and frequency detuning (the difference between the dispersion frequency $\omega_a$ of the mode and the frequency $\omega_0$ at which it oscillates). Importantly, $\overline{\xi}_{a}^{(\mathrm{Sat})}$ scales with a factor $\mathcal{C}$ which is inversely proportional to the strength of the nonlinear self coupling interaction:
\begin{equation}
    \mathcal{C}= \sqrt{\frac{2 c_a}{\omega_0^2 g_{aaaa} + 3 h_{aaaa}}}
\end{equation}
\par
For our experimental parameters, including a theoretical damping rate $\gamma_{4,1}=\SI{0.715\pm0.008}{\per \second}$~\cite{Herreman2019}, Eq.~\ref{eqn:SatAmp} predicts $\overline{\xi}_{4,1}^{(\mathrm{Sat})} =\SI{1.01\pm0.06}{\milli\meter}$ which matches the measured value of $\overline{\xi}_{4,1}(t>35s) = \SI{0.98\pm0.04}{\milli \meter}$ to excellent precision.

\begin{figure*}[ht]
\centering
\includegraphics[width=\linewidth]{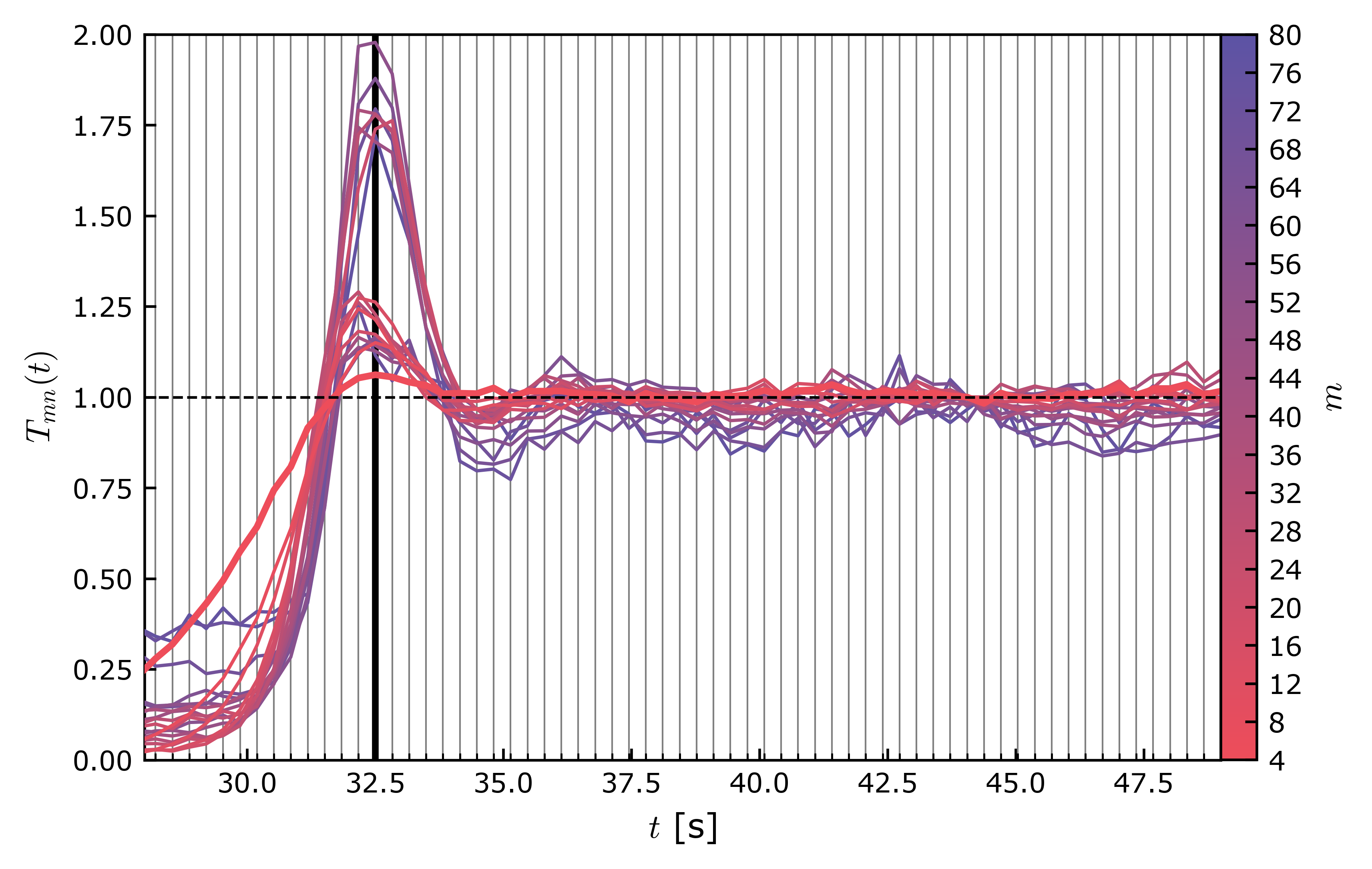}
\caption{\textbf{Transient regime}. To examine the approach towards the steady-state, we examine the root-mean-squared amplitude over one cycle of the primary frequency defined by Eq.~\ref{eq:RMS_amplitude_def}. We define $T_{mn}(t)$ as the ratio between this amplitude at time $t$ and its value $\overline{\xi}^{(\mathrm{Sat})}_{m,n}$ at which it saturates at late times, $T_{mn}(t)\equiv\overline{\xi}_{mn} (t)/\overline{\xi}^{(\mathrm{Sat})}_{m,n}$. This quantity informs us how close the amplitude is to that at which it saturates ($T_{m n } =1$), and has a time resolution up to the primary oscillation cycle $2\pi/\omega_0$ - periods of which have been denoted by grey horizontal lines. We observe that at $\SI{32.5}{\second}$, marked by a bold horizontal line, the primary mode $(4,1)_{\omega_0}$ reaches its maximum amplitude. It then takes 4 cycles of oscillation for it to decay back to its saturation amplitude. Higher $k_{mn}$ modes follow the same behaviour, the transition from their maxima to their saturation amplitude taking around 8 cycles of the the driving frequency $2 \omega_0$ to complete. The full transient regime between nonlinear growth and steady-state scaling occurs over 16 cycles of driving ($\SI{2.58}{\second}$). The modes presented are those excited up to $(m,n)=(80,16)$ only for the sake of visual clarity.}
\label{fig:transient}
\end{figure*}

A reduced dynamics approach is an effective method for calculating aspects of a far-from-equilibrium system undergoing narrow band resonance, and is another context in which our experiment may support cosmological simulation~\cite{bassett1999metric}. 
The timescale for the primary mode to exponentially grow from an initial amplitude $\overline{\xi}_{a}^{(\mathrm{I})}$ to a final amplitude $\overline{\xi}_{a}^{(\mathrm{Sat})}$ is $\Delta t = \lambda_{4,1}^{-1} \log \left(\overline{\xi}_{4,1}^{(\mathrm{Sat})}/\overline{\xi}_{4,1}^{(\mathrm{I})}\right)$. While this is a good approximation for the time until our system reaches a power-law scaled PSD, it does not give the exact timescale to the final steady-state. Instead this gives the timescale until the start of the ``transient regime" during which mode amplitudes transition from exponentially growing to stationary. 
This transient regime commences once the primary mode first attains the amplitude $\overline{\xi}_{4,1}^{(\mathrm{Sat})}$ at which it will saturate. It does not stop growing at this point, but instead overshoots then decays back to this value over the course of eight oscillation cycles. Secondary modes display similar behaviour, fluctuating around their saturation amplitudes $\overline{\xi}^{(\mathrm{Sat})}_b$ during the transient regime before becoming stationary. Extended Data Figure~\ref{fig:transient} depicts the transient regime and subsequent steady state, with each mode amplitude normalised by its ultimate saturation value $\overline{\xi}^{(\mathrm{Sat})}_b$. This feature resembles the behaviour that far-from-equilibrium field theory predicts to occur during the ``damping timescale'' of a systems rapid approach toward a fixed point~\cite{PhysRevLett.93.142002}.

\clearpage
\onecolumngrid

\begin{table} [t]
\centering
\begin{minipage}[b]{\textwidth}
\setlength{\tabcolsep}{18pt}
\renewcommand{\arraystretch}{1.7}
\caption{\textbf{Amplitudes relative to numerical coefficients for the nonlinear interactions $[a.u.]$}. The table displays the numerical value of the wave-amplitudes for node number $n$ in arbitrary units, with the most significant term highlighted in bold, in three different cases: the cubic interaction $(4,1)_{\omega_0}+(4,1)_{\omega_0}\to (8,n')_{2\omega_0}$, and the superposition of the cubic $|A^{(3)}_{12,n''}|$ and quartic $|A^{(4)}_{12,n''}|$ mixing $[(4,1)_{\omega_0}+(8,n')_{2\omega_0}]\bigcup[(4,1)_{\omega_0}+(4,1)_{\omega_0}+(4,1)_{\omega_0}]\to (12,n'')_{3\omega_0}$. The cubic term for this process is significantly lower than the quartic contribution, both peaking at $n=3$. These findings support the experimental observations of the excited modes in the \textit{few} section.}
\label{table:coeff}
\begin{tabular}{  p {3 cm}  p {3 cm}  p {3 cm}  p {3 cm} }
\hline
\hline
$n$ & $|A^{(3)}_{8,n'}|$ & $|A^{(3)}_{12,n''}|$ & $|A^{(4)}_{12,n''}|$\\[1ex]
\hline
0      & 1.2455 & 0.0007 & 1.7366 \\
1      & 0.4040 & 0.0043 & 1.7245 \\ 
2      & \textbf{6.3093} & 0.0163 & 4.9285 \\
3      & 0.3375 & \textbf{0.0431} & \textbf{7.0555} \\
4      & 0.1022 & 0.0168 & 0.5410 \\
5      & 0.0537 & 0.0043 & 0.6574 \\[0.5ex]
\hline
\hline
\end{tabular}
\end{minipage}
\end{table}

\end{document}